\documentclass[11pt]{iopart}
\pdfoutput=1
\expandafter\let\csname equation*\endcsname\relax
\expandafter\let\csname endequation*\endcsname\relax
\usepackage[pdftex]{graphicx,color}
\usepackage{amsmath,amsfonts,amssymb,wasysym,graphicx,capt-of,ifthen,calc}
\usepackage{mathrsfs,graphicx,bbm,amsfonts,amsthm, bm}
\usepackage{enumerate,float,url,color}
\usepackage{amsmath}
\usepackage{graphics}  
\usepackage[latin1]{inputenc}
\usepackage{color}
\usepackage{rotating}
\usepackage{url}
\usepackage{mathtools}
\usepackage{showlabels}
\DeclareMathOperator*{\sgn}{sgn}

 \newcommand{\ba}{\begin{array}}
 \newcommand{\ea}{\end{array}}
 \newcommand{\bea}{\begin{eqnarray}}
 \newcommand{\eea}{\end{eqnarray}}
 \newcommand{\be}{\begin{equation}}
  \newcommand{\ee}{\end{equation}}

 \def \P {{\mathbb P}}
 
 \def \E {{\mathbb E}}

 \def \b {{\beta}}

  \def \g {{\gamma}}

 \def \z {{\z}}

 \def \z {{\zeta}}

\begin{document}
\title{Extreme value statistics of positive recurrent centrally biased random walks}
\author{Roberto Artuso$^{1,2}$, Manuele Onofri$^1$, Gaia Pozzoli$^{1,2}$ and Mattia Radice$^3$}
\address{$^1$Dipartimento di Scienza e Alta Tecnologia and Center for Nonlinear and Complex Systems, Universit\`a degli Studi dell'Insubria, Via Valleggio 11, 22100 Como, Italy}
\address{$^2$I.N.F.N. Sezione di Milano, Via Celoria 16, 20133 Milano, Italy}
\address{$^3$Max Planck Institute for the Physics of Complex Systems, N\"othnitzer Stra\ss e 38, 01187 Dresden, Germany}

\begin{abstract}
We consider the extreme value statistics of centrally-biased random walks with asymptotically-zero drift in the ergodic regime. We fully characterize the asymptotic distribution of the maximum for this class of Markov chains lacking translational invariance, with a particular emphasis on the relation between the time scaling of the expected value of the maximum and the stationary distribution of the process.\\
{\it Keywords:} Stationary distribution, non-homogeneous random walks, extreme value statistics
\end{abstract}

\section{Introduction}
Since their first appearance, stochastic processes have been widely recognized as a fundamental tool for the analysis and prediction of complex systems. It is not surprising that by relying on a probabilistic description based on diffusion theory or random walks, an endless list of phenomena have been explained in physics and a variety of other fields such as biology \cite{Altan}, chemistry \cite{Gardi,Seth} and finance \cite{Gardi,Seth}, to name just a few. One of the most intriguing topics is related to the study of extreme events, e.g. records and maxima, which is covered by the so-called Extreme Value Theory. Such a branch of study finds important applications, for instance, in the analysis of climate phenomena \cite{Bassett,RedPet,MajBomKru}, earthquakes \cite{Sornet} and floods \cite{Bucha,Vogel}, all of which have undoubted relevance in everyday life.

The standard problem consists in deriving the statistics of the maximum of an uncorrelated sequence of $ n $ independent and identically distributed random variables. It is possible to show that in the large-$ n $ regime the limiting distribution is described by one of three possible laws, named Fr\'echet, Gumbel and Weibull, depending on the common probability density function of the individual entries \cite{Fish,Gumbel}. However, such a description is not always the most appropriate and one is often forced to consider some kind of correlation in the sequence, for example when the entries may be interpreted as the steps of a random walk \cite{MajKra,Comtet,MajComZif}. In these cases the problem becomes much more challenging and the emergence of universal laws represents an outstanding result \cite{MajZif,Majumdar-2010}. Interestingly, a few studies have shown that there exists an important connection between the first passage properties and the statistics of extrema, see for example \cite{Bray,Hartich2019b}, implying that for a stochastic process the survival probability and the distribution of the maximum are deeply related.

In this paper we focus on one-dimensional positive recurrent random walks. A random walk is positive recurrent if it returns to the starting point $x_0$ with probability one and finite expected return time. 
Among other reasons, the relevance of this class of processes is due to the fact that usually the statistical analysis of an ergodic process is easier, as will be soon exemplified, and sometimes it is possible to perform proper transformations that may preserve Markovianity but induce stationarity in the modified model: by way of illustration, one can refer to the Lamperti representation for self-affine stochastic processes \cite{LampertiStable1962}, which has recently awakened interest also for experimental purposes \cite {Magdziarz2020}, or consider the edge reinforcement technique for impatient random walks \cite{Englander2019}.
In our context, the recurrence property, together with the further assumptions that each return to the initial position is a renewal event and that the two half-lines of the state space can communicate only through the occurrence of $x_0$, allows us to implement an important simplification. In fact, under these hypotheses, the dynamics can be seen as a sequence of probabilistically identical segments, called \textit{excursions}, which describe the motion between two successive returns. Hence the problem is reduced to the analysis of a sequence of independent and identically distributed random variables, and the distribution of the maximum of the whole process can be written in terms of the maximum of a single excursion. Notably, a finite mean return time guarantees that the law of the maximum, appropriately rescaled, is described in the long time limit by one of the three aforementioned distributions, Fr\'echet, Gumbel or Weibull \textemdash which marks a significant difference with the infinite-mean case \cite{Majumdar-2010,Godr2009,HollBar2020}. Furthermore, we will show that the scaling function, which yields the asymptotic behaviour of the moments and in particular the expected value of the maximum, can be derived from the stationary distribution of the process. 

The rest of the paper is organised as follows. In section \ref{sec:EVT} we summarise the general guidelines on the extreme value analysis for a generic positive recurrent stochastic process. In section \ref{sec:NHRW} we specifically consider some applications to non-homogeneous centrally-biased random walks with asymptotically-zero drift and compare our findings with other results available in the literature. Finally, in section \ref{sec:conclusion} we provide some concluding remarks.

\section{Extreme value statistics of positive recurrent models}\label{sec:EVT}
In the following section, we will consider discrete-time random walks on a lattice, that are sequences of random variables $(X_k)_{k\in\mathbbm{N}_0}$ capable of assuming values in a countable state space $(a_j)_{j\in\mathbbm{Z}}\,$. By performing an appropriate continuum limit, however, in some cases it is possible to carry out limits and recover a continuous model characterized by the same asymptotic statistical features of the original one. 

More precisely, we will be interested in the study of the extreme value statistics of discrete stochastic processes converging to a stationary process: the probability associated with each possible state $a_j$ becomes independent of time and for a fixed set of values $a_{s_1},\dots,a_{s_r}$ the probability
$$
\mathbb{P}[X_{k_1}=a_{s_1},\,X_{k_2}=a_{s_2},\,\dots,\,X_{k_r}=a_{s_r}]
$$
depends only on the differences $|k_i-k_j|\,$. This stationary distribution $(\pi_j)_{j\in\mathbbm{Z}}$ helps to solve the first-passage problem: one can directly find the mean time for which the random process first passes through a specific threshold. Indeed, it is a well known result (Kac Lemma, see Theorem $2$ in \cite{Kac}) that, by taking the inverse of the stationary probability distribution, one immediately obtains also the expected return time to each state $a_j$. In particular, we will consider $\langle \tau\rangle \coloneqq 1/\pi_{X_0}$, which is a key quantity in the characterization of the extreme value statistics, as will soon become clear.

From now on, we will equivalently (in the above meaning) deal with particles diffusing in \emph{sufficiently confining} potentials, in the sense that they ensure positive recurrence and the systems relax to the Boltzmann distributions associated with the potentials. 
The first quantity of interest, as we just said, is the stationary distribution. The Fokker-Planck equation for a one-dimensional diffusive particle subject to a potential $V(x)$ is
\bea
\frac \partial {\partial t} p(x,t)&=&D\frac{\partial^2}{\partial x^2}p(x,t)+\frac \partial {\partial x} \left( \frac d {d x} V(x)p(x,t)\right)\nonumber\\
&=&\frac \partial {\partial x}\left[D\frac \partial {\partial x}+ \frac d {d x} V(x)\right]p(x,t)\,,\label{eq: FP}
\eea
where $p(x,t)$ is the probability density function of the position at a fixed time $t$. Since, by definition, a stationary process does not change with a shift in time, it is straightforward to find that the limiting probability density function $\pi(x)$ is the normalized Boltzmann-Gibbs equilibrium density
\be\label{eq: statDist}
\pi(x)=\lim_{t\to\infty} p(x,t\,|\,x_0)=\frac 1 Z \rme^{-V(x)/D}\,, \qquad \text{for any initial condition }\:x_0\,,
\ee
where $1/Z$ is the normalizing constant.

At this point, after having introduced the two parallel frameworks of interest, it is possible to unveil a less explicit interconnection with the extreme value statistics. The running maximum of a stochastic process $(X_t)_{t\in\mathbbm{R}^+_0}$ up to time $t$ is naturally defined as the largest value assumed by the position in that particular time window, $M_t\coloneqq \max_{s\in[0,t]}X_s$. One way to study its statistics is to deduce information about the maximum of the entire motion from the characterization of a shorter sample between consecutive visits to the starting point, that is called excursion. In the presence of positive recurrence, we know that the mean number of excursions, that are independent of one another due to the renewal property of the process, is estimated by the ratio of the time $t$ over the mean return time $\langle \tau \rangle$. We will consider symmetric potentials with respect to the starting site: as a consequence, for the time being it is possible to assume, for simplicity, that the motion is restricted to the positive semi-axis, with a reflecting barrier at $x_0\,$. Equivalently, it means considering the maximum distance from the starting position on the real line.

Let us denote the maximum of a single excursion by
\be 
M\coloneqq\max_{t\in[0,\tau]}X_t\,,\qquad \tau\coloneqq \min\{t>0\,|\, X_t=x_0\}\,.
\ee
We want to compute
\be
\mathbb{P}(M_t\leq x)=\left[\mathbb{P}(M\leq x)\right]^{\frac t{\langle\tau\rangle}}=[1-Q(x)]^{\frac t{\langle\tau\rangle}}\,,\qquad x\geq x_0\,,
\ee
where $Q(x)$ denotes the splitting probability, that is the probability for a particle diffusing on the finite interval $[x_0,x]$ to eventually hit $x$ without hitting $x_0$, with an appropriate starting position that will be discussed later on. Generally speaking, $Q(x)$ is a function of the initial condition $\bar x\in (x_0,x)$ arbitrarily chosen in the open interval. It is a well known result \cite{Redner} that for isotropic diffusion on a finite interval the first-passage probability to one end-point is just the fractional distance to the other boundary. This suggests that we need another useful mathematical tool, the Lyapunov function \cite{Lyap}, which is an appropriate transformation of the original stochastic process in a confining potential into a motion with no longer drift. Formally, $g:\mathbbm{R}\mapsto \mathbbm{R}$ is the Lyapunov function associated with the process $(X_t)_{t\in\mathbbm{R}^+_0}$ if $Y_t\coloneqq g(X_t)$ is an induced process such that 
\be\label{eq: 0drift}
\mathbb{E}[\Delta_t\,|\, X_t=x]\approx 0\,,
\ee
where $\Delta_t$ represent the increment of the process $Y$ at time $t$, namely
\be
\Delta_t=Y_{t+\delta t}-Y_t=g(X_t+\delta X_t)-g(X_t)= g'(X_t)\cdot \delta X_t+\frac 1 2  g'' (X_t)\cdot \delta X_t^2+\dots\,.
\ee
We recall that 
\be
\dot{X}_t=-\frac {dV(X_t)}{dX_t} +\sqrt{2D}\xi(t)\,,
\ee
which follows from the Langevin equation related to \eqref{eq: FP}, with $\xi(t)$ a $\delta$-correlated gaussian white noise, which means independent at distinct time moments $\langle \xi(t)\xi(t')\rangle=\delta(t-t')$ and characterized by a zero average $\langle \xi (t)\rangle=0\,$. As a consequence, if we perform a discretization and a first-order expansion, thanks to the features of the noise we can write
\bea 
\E[g'(X_t) \delta X_t\,|\, X_t=x]&=&g'(x) \delta t\left(-\frac{dV(x)}{dx}\right)\,,\\
\E[g'' (X_t)\delta X_t^2\,|\, X_t=x]&=&g''(x) \delta t^2\left(\frac {2D}{\delta t}+O(1) \right)\,,
\eea
by observing that $\langle \xi^2(t)\rangle=1/\delta t\,$.
In conclusion, we impose the condition \eqref{eq: 0drift} and get
\be\label{eq: 0driftPot}
-\frac{dV(x)}{dx} g'(x)+D g'' (x)=0\,,
\ee
which identifies the family of functions
\be
g(x)=g_1\int_{x_0}^x\rme^{V(y)/D}\, dy+g_0.
\ee
$g_0$ is an arbitrary constant, since the free motion $(Y_t)_{t\in\mathbbm{R}^+_0}$ is translationally invariant: without loss of generality we can set $g_0=0$. Furthermore, notice that in the absence of a confining potential $g(x)$ must be (up to an eventual translation) the identity function, since the original motion is already a Wiener process: this constraint implies $g_1=1\,$.
At this point, we can therefore compute $Q(x)$ by means of the modified process $(Y_t)_{t\in\mathbbm{R}^+_0}$ and get
\be\label{eq: Q}
Q(x)=\frac{\int_{x_0}^{\bar{x}}\, dy\, \rme^{V(y)/D}}{\int_{x_0}^x\, dy\, \rme^{V(y)/D}}\,.
\ee
An alternative derivation of the result can be found in \cite{MRZ}.

To sum up, keeping in mind \eqref{eq: statDist}, we found that
\be\label{eq: max}
\mathbb{P}(M_t\leq x)=\left[1-\frac {CZ} {\int_{x_0}^x\, dy/\pi(y)}\right]^{\frac t{\langle\tau\rangle}}\,,
\ee
and as a rough estimate it is evident that if there exists a $t$-independent limiting distribution for the maximum, then
\be\label{eq: maxEst}
\int_{x_0}^x\, \frac{dy}{\pi(y)}=O(t)\,.
\ee
This relation directly provides the correct asymptotic behaviour for the maximum in the long-time limit, however one can use a more refined argument. By appropriately shifting and scaling the random variable $M_t$ with respect to $t$, it is also possible to explicitly obtain the exact distribution
\be
\lim_{t\to\infty} \P(M_t\leq a_t+b_tz)=F(z)\,.
\ee
We postpone the details of this last part to the next section, devoted to the study of some specific applications.

\section{Non-homogeneous centrally-biased random walks}\label{sec:NHRW}
From now on we will focus more specifically on the analysis of non-homogeneous discrete-time random walks, which can be traced back to the problem of a particle diffusing in a confining potential by performing an appropriate continuum limit. 

In this framework, an asymptotically-zero drift, whose magnitude tends to zero as the distance from the origin increases, is the weakest possible binding condition that can be imposed in order to ensure positive recurrence. It is known that recurrence properties of a discrete-time irreducible Markov chain $(X_n)_{n\in\mathbbm{N}_0}$ with stationary transition probabilities depend solely on the information carried by the first two moments of the increments
\be 
\mu_k(x)\coloneqq\E[(X_{n+1}-X_n)^k\,|\,X_n=x]\,,\qquad k=1,2\,,
\ee
and they are summarized in the so-called Lamperti criteria \cite{Lamperti1,Lamperti2}. In short, under suitable hypotheses of boundedness about the increments, we know that if
\be\label{eq: lamp}
\limsup_{n\to\infty}\quad [2x\mu_1(x)+\mu_2(x)]\:<\:0\,,
\ee
then the stochastic process is positive recurrent. Uniform boundedness is clearly satisfied by considering nearest-neighbour random walks on $\mathbbm{Z}$, and this will be our case. 

Thus, to summarize, centrally-biased random walks can be seen as perturbations of standard i.i.d. jump processes with zero-drift. The magnitude of the perturbation is the tuning parameter that modifies the asymptotic behaviour of the process, by making the mean return time to the starting position finite. In the following we will deal with an outstanding example, known in the literature as Gillis random walk \cite{Gillis}, which is an emblematic critical stochastic system in the phase transition from null-recurrence to positive-recurrence. Then we will move to some possible generalizations. Our aim, as already stated in the previous section, is to characterize the extreme value statistics and compare it to the moments spectrum, that is the mean asymptotic behaviour of  positive powers of the absolute value of the position with respect to time. In the ergodic regime, there is a marked difference between the moments spectrum and the statistics of the maximum, and this is in contrast to what happens in many examples in the presence of null-recurrence, where the expected maximum and the mean absolute first moment share the same asymptotic growth (and possibly also higher-order moments) \cite{Gillis,Serva,ROAC,Singh}. To provide a heuristic argument, a generic stochastic process on the real line starting from $X_0=0$, symmetric but not necessarily renewal with respect to the initial condition, either is at the maximum distance $\bar M_t\coloneqq \max_{s\in [0,t]}|X_s|$ from the origin where $|X_t|=\bar M_t$ or is confined in the interval $-\bar M_t<X_t<\bar M_t$: in the latter case we say that it experiences off-periods. Notice that, due to the recurrence property and symmetry, it holds that
\be
\langle M_t\rangle \sim K\cdot \langle \bar M_t\rangle\,, \qquad \mbox{with} \quad K\in\left[\frac 1 2,1\right]\,.
\ee
 As a consequence, we just have to focus on the analysis of a single off-period: in the absence of a pretty strong drift toward the origin, it is evident that, since the motion starts from the boundary $\pm \bar M_t$, on average $|X_t|\gtrsim \bar M_t/2$ \cite{Serva}, and one can therefore conclude that $\langle |X_t|\rangle$ is of the same order of magnitude of $ \langle M_t\rangle$:
\be
\langle |X_t|\rangle \asymp \langle M_t\rangle\,,\qquad \mbox{as}\quad t \to \infty.
\ee
This effect can be possibly enhanced by the presence of long ballistic flights dominating the dynamics, in the course of which there is an equivalence between position and maximum\textemdash see also the big jump principle \cite{VeBaBu-2019,VeBaBu-2020,Holl-2021}. On the other hand, the confining action of a significant bias, responsible for the existence of a stationary distribution, can instead break up this correspondence, during the off-periods, between the absolute value of the position and the maximum distance reached by the process up to that time.
 
\subsection{Gillis random walk in the ergodic regime.}
The Gillis model is one of the few analytically solvable non-homogeneous random walks whose key defining feature is a drift dependent on the position in the sample. The lack of translational invariance is clear from the definition of the transition probabilities, since there appears an explicit dependence on the current site. The walker starts at the origin $X_0=0$ and then moves on $\mathbbm{Z}$ according to the following rules: if $p_{i,j}$ denotes the probability of moving from site $i$ to site $j$, we have
\be\label{eq: GillisDef}
\mathcal{R}(j)\coloneqq p_{j,j+1}\,,\quad \mathcal{L}(j)\coloneqq p_{j,j-1}\,,\qquad p_{i,j}=0\quad \mbox{if}\quad |i-j|\neq 1\,,
\ee
where
\be
 p_{j,j\pm1}\coloneqq\frac 1 2 \left(1\mp \frac \epsilon j\right)\quad \mbox{for}\quad j\in\mathbbm{Z}\backslash\{0\}\,,\quad \mathcal{R}(0)\coloneqq \frac 1 2 \eqqcolon \mathcal{L}(0)\,,
\ee
and the real parameter $\epsilon\in(-1,1)$ tunes the bias toward or away from the origin. From \eqref{eq: lamp} it is immediate to verify that in order to guarantee positive-recurrence it is necessary to require that $\epsilon\in\left(\frac 1 2, 1\right)$. In fact, we have
\be
\mu_1(j)=\mathcal{R}(j)-\mathcal{L}(j)=-\frac{\epsilon}{j}\,,\qquad \mu_2(j)=\mathcal{R}(j)+\mathcal{L}(j)=1\,.
\ee
Positive recurrence, together with aperiodicity and irreducibility (that implies ergodicity) of the Markov chain, allows us to state that there exists a unique stationary distribution $\pi_j\,$, with $\pi_{-j}=\pi_j$ by symmetry. More explicitly, this is defined by
\be
\pi_j=\sum_{i}p_{i,j}\pi_i,\qquad\mbox{with}\quad\sum_{j}\pi_j=1,
\ee
which leads to a simple expression written in terms of the transition probabilities $ \mathcal{L}(j) $ and $ \mathcal{R}(j) $ and the probability $ \pi_0 $ of being at $ j=0 $:
\be\label{eq: Stat_general}
\pi_j=\pi_0\begin{cases}
	\displaystyle\frac 1 {2\mathcal{L}(1)}\qquad&\mbox{if}\quad |j|=1\,,\\
	\displaystyle\frac 1 {2\mathcal{L}(|j|)} \prod_{k=1}^{|j|-1}\frac{\mathcal{R}(k)}{\mathcal{L}(k)}\qquad &\mbox{if}\quad |j|>1.\\
\end{cases}
\ee
For the Gillis model, with a simple computation it is possible to fully determine the stationary distribution \cite{Onofri}, which remarkably shows an asymptotic power-law decay
\be \label{eq: stat}
\pi_j=\frac{|j|(1-\epsilon)_{|j|-1}}{(1+\epsilon)_{|j|}}\pi_0\sim \pi_0\frac{\Gamma(1+\epsilon)}{\Gamma(1-\epsilon)}|j|^{-2\epsilon}\qquad \mbox{as}\quad j\to\pm\infty\,,
\ee
where $(x)_k$ is the Pochhammer symbol (or rising factorial) and 
\be\label{eq:Bir}
\pi_0=\frac 1 {\langle\tau\rangle}=\frac{2\epsilon-1}{2\epsilon}\,,
\ee
according to the Birkhoff ergodic theorem. In fact, it is easy to observe that in the long-time limit the time average of the number of returns to the origin $R_n=\sum_{k=1}^n \delta_{X_k,0}$ converges to the ensemble average of the Kronecker delta, which is $\pi_0$ by definition, and at the same time $\langle \tau\rangle $ is trivially given by the ratio of the number of steps over the mean number of returns $R_n$. Thus \eqref{eq:Bir} immediately follows, by referring to \cite{Hughes} for the computation of the expected return time.

As  we emphasized in our previous works \cite{Onofri,Pozzoli}, this random walk is a discrete realization of a diffusing particle in the presence of an asymptotically logarithmic potential. Indeed, if one considers the master equation governing the time evolution of the probability $p(j,n)$ of finding the walker at site $j$ after $n$ steps
\be
p(j,n+1)=p(j-1,n)\mathcal{R}(j-1)+p(j+1,n)\mathcal{L}(j+1)\,,
\ee
and defines a lattice spacing $\delta x$ and a time increment $\delta t$ in such a way that $\delta x,\delta t\to 0\,$, $x=j\delta x$, $t=n\delta t$ together with the diffusion approximation $\delta x^2/\delta t\to D_0$ (set equal to $1$ without loss of generality), one gets the Fokker-Planck equation
\be \label{eq: GillisCL}
\frac \partial {\partial t} p(x,t)=\frac 1 2\frac{\partial^2}{\partial x^2}p(x,t)+\frac \partial {\partial x} \left( \frac \epsilon xp(x,t)\right)\,.
\ee
Actually, in a neighbourhood of the origin, denoted by the interval $(-a,a)$, the particle diffuses freely, as can be seen from the definition of the model in \eqref{eq: GillisDef}, and so, heuristically, the potential obtained in \eqref{eq: GillisCL} must be regularized. More correctly, with a simple argument (see \ref{app: gillis} for a detailed discussion), one obtains that 
\begin{equation}
	V(x)=
	\begin{dcases}
		0 &\quad \text{for }\:|x|<a\,,\\
		\frac 1 2\log\left[\frac{\Gamma(1-\epsilon)}{\Gamma(1+\epsilon)}\frac{\Gamma(|x|+1+\epsilon)}{|x|\Gamma(|x|-\epsilon)}\right] &\quad \text{for }\:|x|>a\,,
	\end{dcases}\label{eq: potGillisCL}
\end{equation}
where $V(x)\sim\epsilon \log (|x|)$ for $x\gg 1\,$, according to Eq. \eqref{eq: GillisCL}.
Notice that the length of the interval of free diffusion is a crucial quantity to ensure the validity of the continuum limit \eqref{eq: potGillisCL} and can not be chosen arbitrarily.  More precisely, the parameter $a$ must be fixed by imposing to obtain matching results with respect to those of the original discrete model, and in particular one can exploit the consistency check for the stationary distribution \eqref{eq: stat}: a detailed discussion is postponed to \ref{app: gillis}. For our purposes, it is sufficient to observe that at large distances from the origin
\be \label{eq: pot}
V(x)\sim \epsilon\log\left( \frac {|x|}{\eta}\right)\qquad \mbox{with}\quad 
 {\eta^{2\epsilon}}=\frac{\Gamma(1+\epsilon)}{\Gamma(1-\epsilon)}\,,
\ee
or equivalently 
\be\label{eq: potStat}
\pi(x)\sim \frac 1 Z \left(\frac \eta x \right)^{2\epsilon}\,, \qquad \mbox{with}\quad Z\equiv \frac 1 {\pi_0}=\frac{2\epsilon}{2\epsilon-1}\,.
\ee
As a final comment, it is worth noting that the regularization of the potential at the origin corresponds to similarly consider the identity as Lyapunov function in the interval of free diffusion and combine it with the non-trivial transformation at large distances from the origin.

At this point, we are therefore equipped to move on to the analysis of the extreme value statistics. 
In order to deal with a symmetric random walk $Y_n\coloneqq g(X_n)$ with no drift, as explained above we can take a test function $g(x)=x^\gamma$ that inserted in \eqref{eq: 0drift} gives
\bea
\mathbb{E}[\Delta_n\,|\, X_n=x]&=&\E[Y_{n+1}-Y_n\,|\,X_n=x]=\E[X_{n+1}^\g-X_n^\g\,|\,X_n=x]\nonumber\\
&=&\E\left[ X_n^\g\left[\left(1+\frac{X_{n+1}-X_n}{X_n}\right)^\g-1\right]\,\Big|\,X_n=x\right]\nonumber\\
&=&\g\mu_1(x)x^{\g-1}+\frac{\g(\g-1)}2\mu_2(x)x^{\g-2}+\dots\nonumber\\
&=& \g\left(-\epsilon+\frac{\g-1}2\right)x^{\g-2}+\dots\approx 0\,,\qquad \forall x\,,
\eea
which results in $\g=2\epsilon+1$. Alternatively, the Lyapunov function can be obtained as the solution of (see \eqref{eq: 0driftPot})
\be
g'(x)\left(-\frac {2\epsilon} x\right)+g''(x)=0\,, \qquad \mbox{for}\quad |x|\gg a\,.
\ee
To be more accurate and determine the scaling as well as the constants of proportionality, we can directly exploit the stationary distribution \eqref{eq: potStat} associated with the logarithmic confining potential \eqref{eq: pot} to be replaced in \eqref{eq: max} and write
\be
\mathbb{P}(M_t\leq x)\sim \left[1-\frac {\eta^{2\epsilon}(2\epsilon+1)}{x^{2\epsilon+1}}\right]^{\frac t{2\langle\tau\rangle}}\qquad\mbox{as}\quad t,x\gg 1\,,
\ee
where the exponent is the number of excursions divided by two since by symmetry in half of the cases the walker explores the negative semi-axis and the maximum stays below the threshold $x>0$ with probability one.
Notice that we set $C\sim 1$ in \eqref{eq: max} since, in the discrete-time random walk on the lattice, the excursion begins when the walker jumps to the right (or to the left) of the origin and this determines the correct fractional distance from the origin for the modified process with no drift \cite{Redner}. For further details, refer to \ref{app: gillis}.

\begin{figure}[htbp]
\centering
\includegraphics[width=0.48\textwidth]{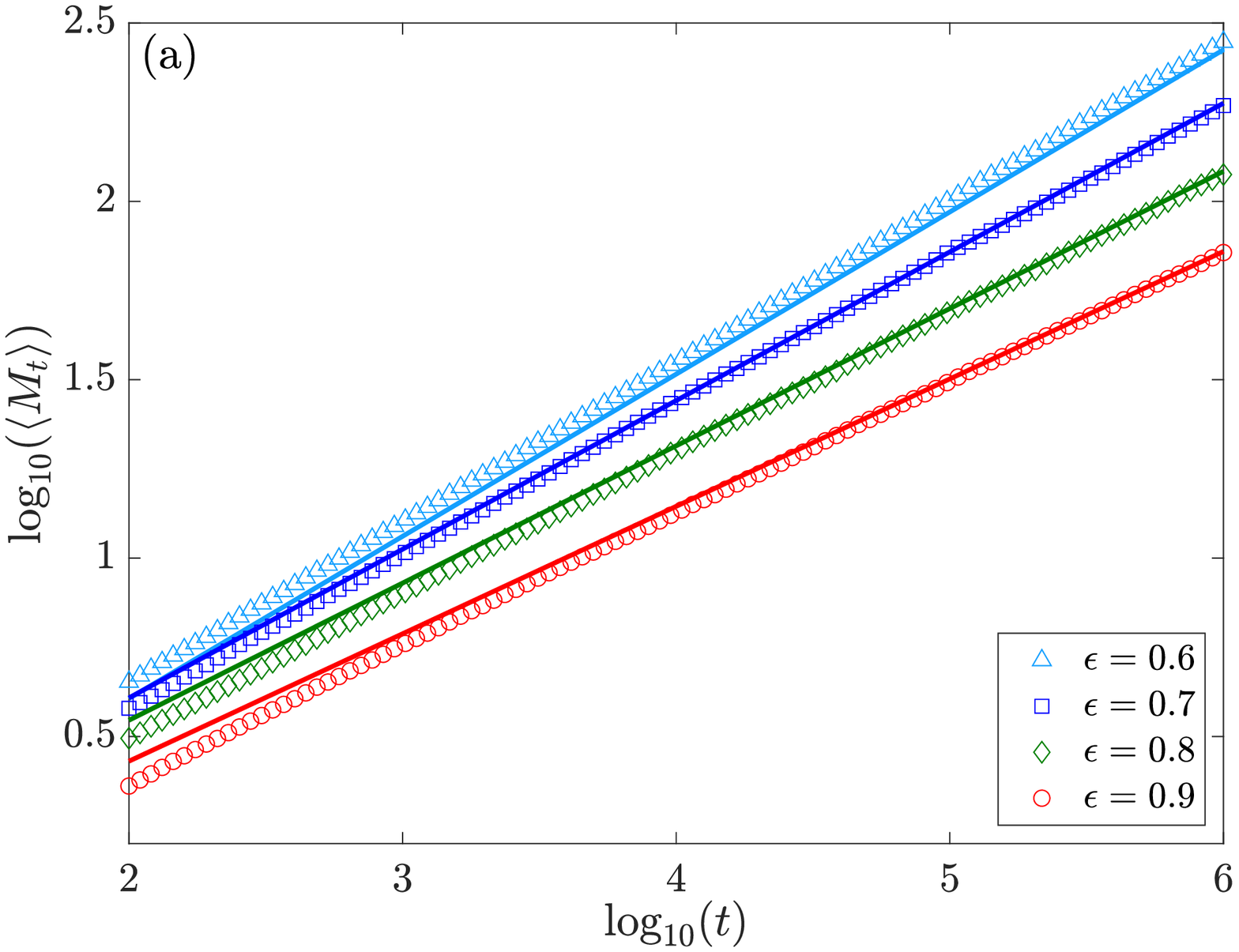}
\includegraphics[width=0.48\textwidth]{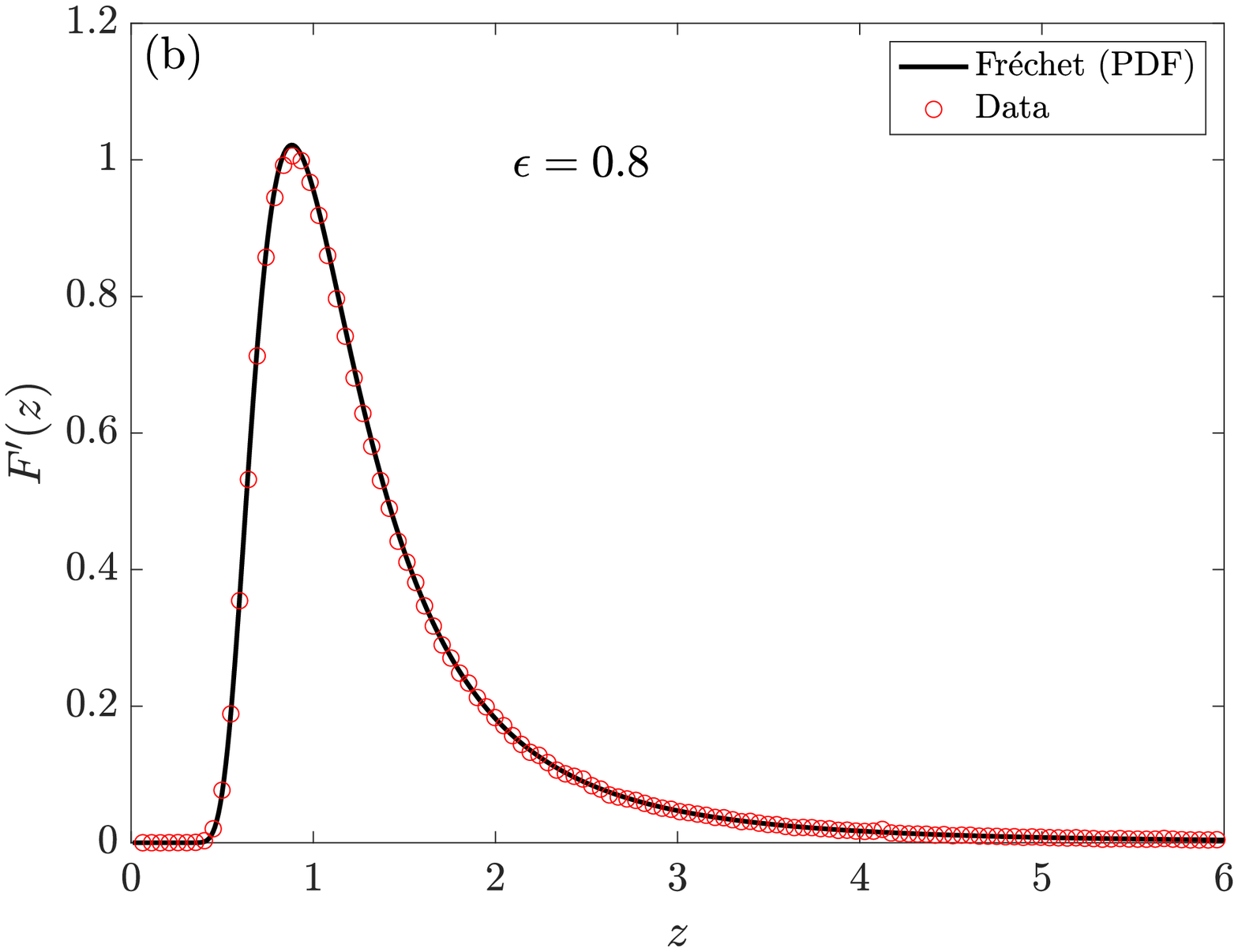}
\caption{(a) Expected maximum as a function of time: markers represent data obtained simulating $10^4$ walks up to time $10^6$ for different values of $\epsilon$, lines are the corresponding theoretical predictions given by \eref{eq:Max_Gillis}; (b) Fr\'echet distribution for $10^6$ walks evolved up to time $10^6$ with $\epsilon=0.8$.}
\label{fig: gillis erg}
\end{figure}

It is easy to observe that the change of variable $x=b_t z\,$, with 
$$b_t=\left(\frac {\eta^{2\epsilon}(2\epsilon+1)}{2\langle \tau\rangle}t\right)^{1/(2\epsilon+1)}\,,$$
leads to a Fr\'echet distribution
\be
F(z)=\rme^{-z^{-2\epsilon-1}}\,,\qquad \mbox{with mean}\quad \Gamma\left(1-\frac 1 {2\epsilon+1}\right)\,.
\ee
In particular, we can therefore conclude that in the long-time limit the asymptotic behaviour of the expected maximum is completely determined by
\be\label{eq:Max_Gillis}
\langle M_t\rangle \sim \left[\frac {\eta^{2\epsilon}(2\epsilon+1)}{2\langle \tau\rangle}\right]^{1/(2\epsilon+1)}\Gamma\left(1-\frac 1 {2\epsilon+1}\right)\cdot t^{1/(2\epsilon+1)}\,.
\ee
For a numerical check, see figure \ref{fig: gillis erg}.

\subsection{Gillis persistent random walk in the ergodic regime.} 
At this point, one can wonder what happens if we define transmission and reflection coefficients dependent on the position on the one-dimensional lattice, which means to inquire about the potentiality of non-homogeneous persistence. A first non-Markovian generalization of Lamperti criteria can be found in \cite{georgiu}, concerning random walks with transmission and reflection coefficients of the form
\be
t_j^\pm\coloneqq\frac{1}{2}\left(1-\frac{\epsilon_\pm}{|j|}\right)\,, \qquad r_j^\pm\coloneqq\frac{1}{2}\left(1+\frac{\epsilon_\pm}{|j|}\right)\quad\mbox{for}\quad j\neq 0\,,\qquad t^\pm_0\equiv r^\pm_0\coloneqq\frac 1 2\,,
\ee
where $+$ denotes the outward direction, whereas $-$ the inward motion with respect to the origin.
It is easy to show (see \ref{app: gillis pers}) that, in the continuum limit, these processes are equivalent to a Gillis random walk with parameter $\epsilon=\frac{\epsilon_+-\epsilon_-}{2}\,$. As a consequence, the generalized theorem characterizing the recurrence properties (see Corollary $3.1$ in \cite{georgiu}) can be directly derived from the application of the original version of Lamperti criteria and one can conclude that positive-recurrence is ensured whenever $\epsilon_+-\epsilon_->1$. It is worth noting that in the symmetric case ($\epsilon_+=\epsilon_-$) we have free diffusion: diversification in the coupling with the two possible directions of motion is therefore essential in the persistent model.

The most general non-homogeneous persistent random walk, however, stems from the perturbation of a homogeneous Markov chain with constant transmission and reflection coefficients $p\neq q\left(\neq\frac{1}{2}\right)$ \cite{gillis_corr,renshaw}. For a general introduction to one-dimensional correlated random walks having diverse intrinsic probabilities in different directions refer to \cite{domb, seth, chen}.
In our context, we can explicitly define
\begin{equation}\label{eq: persGen}
t^{\pm}_j\coloneqq q-\frac{\epsilon_{\pm}}{2|j|}\,,\qquad r^{\pm}_j\coloneqq 1-q+\frac{\epsilon_{\pm}}{2|j|}\,,\qquad q\in(0,1)\,,
\end{equation}
with a proper regularization for small $|j|$. This is asymptotically indistinguishable from a standard Gillis random walk with $\epsilon=\frac{\epsilon_+-\epsilon_-}{4q}\,$, given that the diffusion equation in the continuum limit becomes
\begin{equation}\label{eq: persGenCL}
\frac{\partial}{\partial t} p(x,t)=\frac {D_0} 2\frac{q}{1-q}\left[\frac{\partial ^2}{\partial x^2}+\frac{\epsilon_+-\epsilon_-}{2q}\frac{\partial}{\partial x}\left(\frac{1}{x}p(x,t)\right)\right]\,.
\end{equation}
Therefore, by applying once again Lamperti criteria, we know that there exists a unique stationary distribution if and only if $\epsilon_+-\epsilon_->2q\,$.
In particular, if $q=\tfrac{1}{2}$, we recover the statement of Corollary $3.1$ in \cite{georgiu}. For a rigorous analysis of the generalized model $q\neq \tfrac 1 2$, instead, refer to Theorem $2.6$ in \cite{wade} and \ref{app: gillis pers}. 

Turning back to our study of the maximum statistics, it is now clear that the asymptotic scaling with respect to the time determined in the previous subsection remains valid also in these examples, by replacing in a proper way all the parameters.

\subsection{Generalized Gillis model: supercritical power-law drift.}\label{subsec:GGL}
Now we can think to examine also the supercritical cases, where $|x\mu_1(x)|\to\infty$. Starting from \eqref{eq: GillisDef}, we can slightly change the definition of the transition probabilities by introducing an additional parameter $\beta\in(0,1)$ such that
\be
 p_{j,j\pm1}\coloneqq\frac 1 2 \left(1\mp \sgn(j) \frac \epsilon {|j|^\beta}\right)\quad \mbox{for}\quad j\neq 0\,, \qquad \mu_1(j)=-\sgn(j)\epsilon/|j|^\beta\,.
\ee
Clearly, the mean first-return time to the origin is finite if and only if $\epsilon>0$. Similarly to the Gillis model in the ergodic regime, by using \eref{eq: Stat_general} it is possible to define recursively the discrete stationary distribution
\be\label{eq: superStat}
\pi_j=\pi_0\begin{cases}
\displaystyle\frac 1 {1+\epsilon}\qquad &\mbox{if}\quad |j|=1\,,\\
\displaystyle\frac{|j|^\beta}{|j|^\beta+\epsilon}\prod_{k=1}^{|j|-1}\frac{k^\beta-\epsilon}{k^\beta+\epsilon}\qquad&\mbox{if}\quad |j|>1\,,\\
\end{cases}
\ee
even though we can not carry out an explicit computation as in \eqref{eq: stat}. Anyway, for our aim we are only interested in the long-distance asymptotic behaviour, which can be extracted by first observing that for $ |j|>1 $ we may write
\begin{equation}\label{eq:super_atan}
	\pi_j=\pi_0\exp\left[-2\sum_{k=1}^{|j|-1}\mathrm{atanh}\left(\frac{\epsilon}{k^\beta}\right)-\log\left(1+\frac{\epsilon}{|j|^\beta}\right)\right],
\end{equation}
where $ \mathrm{atanh}(z) $ is the inverse hyperbolic tangent and we used the relation \cite{Abr-Steg}
\begin{equation}
	2\mathrm{atanh}(z)=\log\left(\frac{1+z}{1-z}\right),\qquad0\leq z^2<1.
\end{equation}
To estimate the summation in \eref{eq:super_atan}, we replace $ \mathrm{atanh}(z) $ with its MacLaurin series, obtaining
\begin{align}
	\sum_{k=1}^{|j|-1}\mathrm{atanh}\left(\frac{\epsilon}{k^\beta}\right)&=\sum_{k=1}^{|j|-1}\sum_{n=0}^{\infty}\frac{\epsilon^{2n+1}}{2n+1}k^{-(2n+1)\beta}\nonumber\\
	&=\sum_{n=0}^{\infty}\frac{\epsilon^{2n+1}}{2n+1}\sum_{k=1}^{|j|-1}k^{-(2n+1)\beta}.
\end{align}
We now observe that in the limit of large $ |j| $ the leading order term is obtained for $ n=0 $. By approximating the sum over $ k $ with an integral we get
\begin{equation}\label{eq:Pot_super_est}
	\sum_{k=1}^{|j|-1}\mathrm{atanh}\left(\frac{\epsilon}{k^\beta}\right)\approx\epsilon\int_{1}^{|j|-1}dk\,k^{-\beta}=\frac{\epsilon}{1-\beta}\left[(|j|-1)^{1-\beta}-1\right],	
\end{equation}
hence by retaining only the leading term for $ |j|\gg1 $ in the above expression, the resulting asymptotic behaviour of $ \pi_j $ is given by
\begin{equation}
	\pi_j\sim\pi_0\exp\left(-\frac{2\epsilon|j|^{1-\beta}}{1-\beta}-\eta\right).
\end{equation}
The validity of this estimate can be confirmed by considering the continuum limit. It is easy to prove that this random walk is the discrete version of a diffusing particle subject to a force asymptotically characterized by a power-law dependence on the distance from the origin. 
Indeed, by Taylor expanding the master equation, one gets
\be 
\frac \partial {\partial t} p(x,t)=\frac{\delta x^2}{\delta t}\left[ \frac 1 2\frac{\partial^2}{\partial x^2}p(x,t)+\sgn(x)\epsilon\delta x^{\beta-1}\frac \partial {\partial x} \left( \frac {p(x,t)}{|x|^\beta}\right)\right]\,,
\ee
and by requiring, in the continuum limit $\delta x,\delta t\to 0$, that $\epsilon\to 0$ and the product $\varepsilon\coloneqq \epsilon\delta x^{\beta-1}$ remains finite, the diffusion equation becomes
\be 
\frac \partial {\partial t} p(x,t)=\frac 1 2\frac{\partial^2}{\partial x^2}p(x,t)+\sgn(x)\frac \partial {\partial x} \left( \frac \varepsilon {|x|^\beta}p(x,t)\right)\,,
\ee
that implies handling with a potential of the form
\be
V(x)\sim\frac \varepsilon {1-\beta}|x|^{1-\beta}\,,\qquad \mbox{as}\quad |x|\gg1\,,
\ee
in agreement with \eref{eq:Pot_super_est} by setting $\varepsilon\equiv\epsilon\,$. Hence, turning to the asymptotic behaviour of the maximum, in \eqref{eq: max} we can consider the continuum approximation of the stationary distribution
\be\label{eq: statApprox}
\pi(x)\sim\frac 1 Z \exp\left(-\frac{2\epsilon|x|^{1-\beta}}{1-\beta}-\eta\right)\,.
\ee
Let us point out that $Z$ and $\eta$ are no longer explicit, but we anticipate that they will not take an active role in our estimates. Moreover, we postpone to \ref{app:StatSuperGillis} a numerical analysis on the validity of \eqref{eq: statApprox}.
Before proceeding further, notice that all the moments converge to a constant, due to the stretched exponential form of the stationary solution, whereas in the original model the slow decay $|x|^{-2\epsilon}$ makes the high-order moments to diverge, providing non-trivial transport properties. 

As a first step, starting from \eqref{eq: max} we need to compute the integral
\begin{align}
\int_0^x\, dy\, \rme^{\lambda y^{1-\beta}}&=\frac 1 \lambda \int_0^x \, dy\, \frac{y^\b}{1-\beta}\, \frac d {dy} \rme^{\lambda y^{1-\beta}}=\frac{x^\b \rme^{\lambda x^{1-\beta}}}{\lambda(1-\beta)} -\frac 1 \lambda \int_0^x\, dy\, \frac{\rme^{\lambda y^{1-\beta}}}{y^{1-\b}}\frac \beta {1-\beta}\nonumber\\
&\sim\frac{x^\b \rme^{\lambda x^{1-\beta}}}{\lambda(1-\beta)} +\dots
\end{align}
 as $x\to \infty\,$. Hence we can write
\be\label{eq: maxSuper}
\mathbb{P}(M_t\leq x)\sim \left[1-\frac {2 C\epsilon  \rme ^{-\eta}} {x^\b \rme^{\frac{2\epsilon}{1-\b}x^{1-\b}}}\right]^{\frac t{2\langle\tau\rangle}}\,,
\ee
and impose the condition \eqref{eq: maxEst} in order to obtain the correct scaling with respect to $t$, for $t\gg 1\,$. What we get is an equation where the unknown quantity appears both in the base and the exponent, and so it can be solved by using the Lambert $W$ function, which is defined as the inverse function of $f(z)=z\rme^z\,$ with $z$ any complex number. By means of the change of variable $w\coloneqq x^{1-\b}\,$, our equation can be converted into an equation of the form
\be\label{eq: lambert}
A^w=B\cdot w^E\,,\qquad \mbox{with}\quad \begin{cases}
A=\rme^{\frac{2\epsilon}{1-\b}}\\
\displaystyle B=\frac{C\epsilon t}{\rme ^\eta\langle \tau\rangle}\\
E=-\frac \b {1-\b}\\
\end{cases}\,,
\ee
and it is not difficult to show that the solution is
\be 
w=-\frac E {\log(A)}W\left(-\frac{\log(A)}{EB^{1/E}}\right)\,.
\ee
In conclusion, by substituting our parameters, we have
\be \label{eq: meanMaxSuperRough}
x=\left[\frac\b {2\epsilon} W\left( \frac{2\epsilon}\b\left(\frac{C\epsilon t}{\rme ^\eta\langle \tau\rangle} \right)^{\frac 1 \b -1}\right) \right]^{\frac 1 {1-\b}}\sim \left(\frac{1-\b}{2\epsilon}\log(t)\right)^{\frac 1 {1-\b}}\qquad \mbox{as}\quad t \to \infty\,.
\ee
It is worthwhile to notice that the pairing of a stretched exponential with a power of the logarithm is analogous to the relationship between the exponential stationary distribution and the logarithmic growth of the expected maximum for stochastic processes with Poissonian resetting \cite{Godreche2022} (see also section \ref{sec:overview}).

\begin{figure}[htbp]
	\centering
	\includegraphics[width=0.48\textwidth]{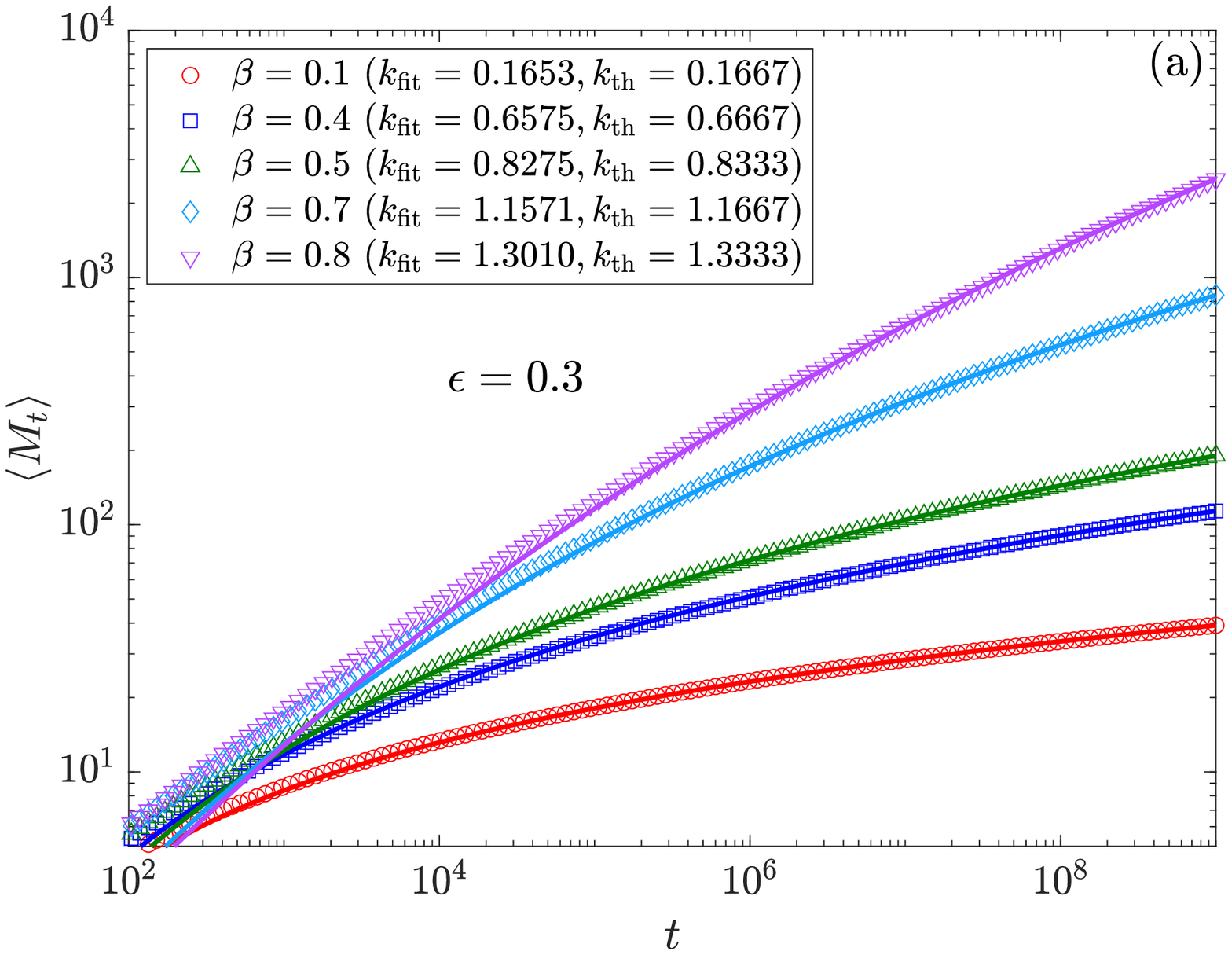}
	\includegraphics[width=0.48\textwidth]{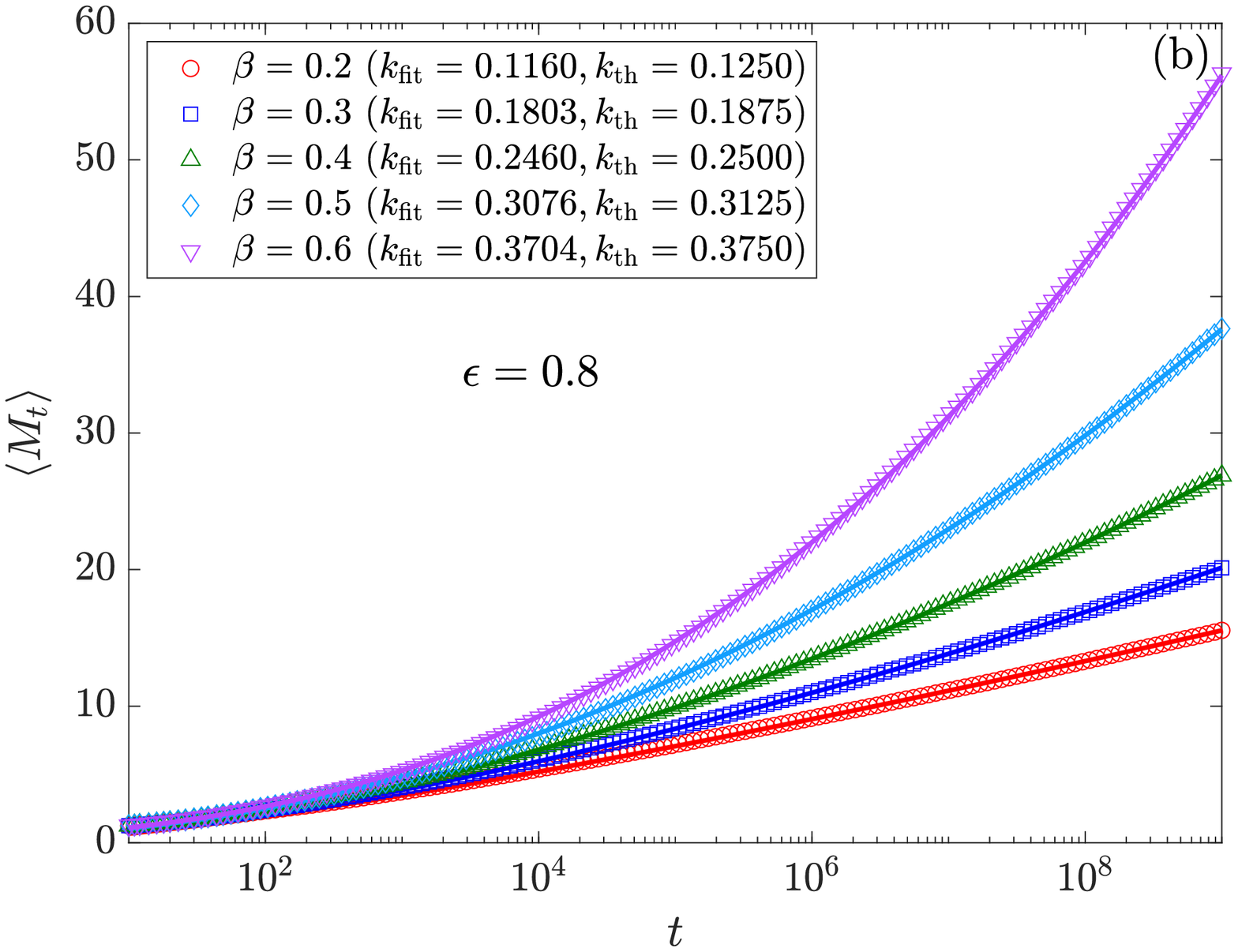}
	\caption{Expected maximum as a function of time for the supercritical model for several values of $ \beta $, with (a) $ \epsilon=0.3 $ and (b) $\epsilon=0.8$. In both panels, the markers represent data obtained simulating $10^4$ walks up to time $10^9$. The solid lines are fitted curves, using the expression of $ a_t $ given by \eqref{eq: scalSuper}. For each fixed value of $ \beta $, the fit function used is of the kind $ y=\left[kW\left(qx^{1/\beta-1}\right)\right]^{1/(1-\beta)} $, with $ k $ and $ q $ left as free parameters. In the plot are reported the fitted $ k $ and the corresponding theoretical values $ k_{\mathrm{th}}=\frac\beta{2\epsilon} $.}
	\label{fig: gillis super}
\end{figure}

For a more accurate analysis, we have to find the transformation $x=a_t+b_tz$ that ensures the convergence to a Gumbel distribution
\be
\lim_{t\to\infty} \P(M_t\leq a_t+b_tz)=\rme^{-\rme^{-z}}\,,
\ee
which is expected since the decay in \eqref{eq: maxSuper} is slower than a pure exponential. By comparison with \eqref{eq: maxSuper}, this is clearly equivalent to requiring that
\be
\lim_{t\to\infty}\frac{C\epsilon t}{\rme^\eta\langle\tau\rangle} \frac{\rme^{-\frac{2\epsilon}{1-\b}(a_t+b_tz)^{1-\b}}}{(a_t+b_tz)^\b}=\rme^{-z}\,.
\ee
Thus we have to perform a series expansion of the left-hand side 
\be
\lim_{t\to\infty} \frac{C\epsilon t}{\rme ^\eta \langle\tau\rangle} \frac{\rme^{-\frac{2\epsilon}{1-\beta}a_t^{1-\beta}\left(1+\frac{b_t}{a_t}(1-\beta)z+\dots\right)}}{a_t^\beta\left( 1+\frac{b_t}{a_t}\beta z+\dots\right)}\,,
\ee
and compare the coefficients of the powers $z^0$ and $z^1$ on both sides
\begin{subequations}
\begin{eqnarray}
\rme^{\frac{2\epsilon}{1-\beta}a_t^{1-\beta}}&\sim &\:\frac{C\epsilon t}{\rme ^\eta \langle\tau\rangle} a_t^{-\beta}\,,\\
 \rme^{-z}&\sim &\: \rme^{-2\epsilon a_t^{1-\beta}\frac{b_t}{a_t}z}\rme ^{-\beta \frac{b_t}{a_t}z}\,,
\end{eqnarray}
\end{subequations}
 to conclude that
\be\label{eq: scalSuper}
a_t=\left[\frac\b {2\epsilon} W\left( \frac{2\epsilon}\b\left(\frac{C\epsilon t}{\rme^\eta\langle \tau\rangle} \right)^{\frac 1 \b -1}\right) \right]^{\frac 1 {1-\b}}\,,\qquad  b_t=\frac{a_t}{2\epsilon a_t^{1-\beta}+\beta}.
\ee
Since the mean value of the Gumbel distribution is the Euler-Mascheroni constant $\g\,$, we can affirm that the expected maximum in the long-time limit is given by
\be
\langle M_t\rangle \sim a_t+b_t\g\sim a_t+\frac{\g}{2\epsilon} a_t^\beta\qquad\mbox{as }t\to\infty\,,
\ee
see figure \ref{fig: gillis super} for a numerical check; thus by using the asymptotic expansion of the principal branch of the Lambert $ W $ function we obtain that up to leading order
\begin{equation}\label{key}
	\langle M_t\rangle\sim \left(\frac{1-\b}{2\epsilon}\log(t)\right)^{\frac 1 {1-\b}},
\end{equation} 
which confirms the scaling we found with the rough estimate \eqref{eq: meanMaxSuperRough}.

\subsubsection{The limit case $ \beta=0 $}
It is instructive to study the case where the bias is still directed toward the origin, but its intensity does not depend on the distance from it. In this way, clearly, the asymptotically-zero drift property breaks down. This corresponds to setting $ \beta=0 $ in the previous supercritical model. From a physical point of view, the motion corresponds to the diffusion of a Brownian particle in a constant central field. In this simple scenario we can arrive to an explicit expression for the stationary distribution, which reads
\be\label{eq:beta0Stat}
\pi_j=\pi_0\begin{cases}
	\displaystyle1\qquad &\mbox{if}\quad j=0\,,\\
	\displaystyle\frac1{1-\epsilon}\left(\frac{1-\epsilon}{1+\epsilon}\right)^{|j|}\qquad&\mbox{if}\quad |j|\geq1.\\
\end{cases}
\ee
The value of $ \pi_0 $ can be determined from the normalization condition, yielding $ \pi_0=\epsilon/(1+\epsilon) $. Note that, except for the point $ j=0 $, we have $ \pi_j =\pi_0\rme^{-\lambda|j|-\eta}$, with
\be
\lambda=2\mathrm{atanh}(\epsilon),\qquad\eta=\log(1-\epsilon),
\ee
therefore we can estimate the distribution of the maximum from \eref{eq: max} by using the continuum approximation (see also \ref{app: gillis})
\be\label{eq: beta0StatCont}
\pi(x)=\frac{\rme^{-\lambda|x|-\eta}}{Z}.
\ee
We get
\be
\mathbb{P}\left(M_t\leq x\right)\sim\left(1-\frac{C\lambda\rme^{-\eta}}{\rme^{\lambda x}}\right)^{\frac t{2\langle\tau\rangle}},
\ee
where $ \langle\tau\rangle $ can be replaced by its explicit expression given by
\be
\langle\tau\rangle=\frac1{\pi_0}=\frac{1+\epsilon}\epsilon.
\ee
It is not difficult to obtain the transformation $ x=a_t+b_tz $ leading to a Gumbel distribution, described by the coefficients
\be
a_t=\frac1\lambda\log\left(\tilde{t}\right),\qquad b_t=\frac1\lambda\,,
\ee
where $ \tilde{t} $ is the rescaled time
\be
\tilde{t}=\frac{\lambda C}{2\rme^{\eta}}\frac{t}{\langle\tau\rangle},
\ee
and thus in the long-time limit we observe a logarithmic asymptotic growth of the expected maximum
\be\label{eq:beta0_Max}
\langle M_t\rangle\sim\frac1{\lambda}\log(t).
\ee
Interestingly, the same law describes the asymptotic behaviour of $ \langle M_t\rangle $ for a random walk with stochastic resetting \cite{Majumdar2022,Godreche2022}, which indeed exhibits a stationary state of the same form of our $ \pi(x) $, see \eref{eq: beta0StatCont}. This reinforces our observation regarding the relation between the stationary distribution and the limiting distribution of the maximum.
\begin{figure}[htbp]
	\centering
	\includegraphics[width=0.48\textwidth]{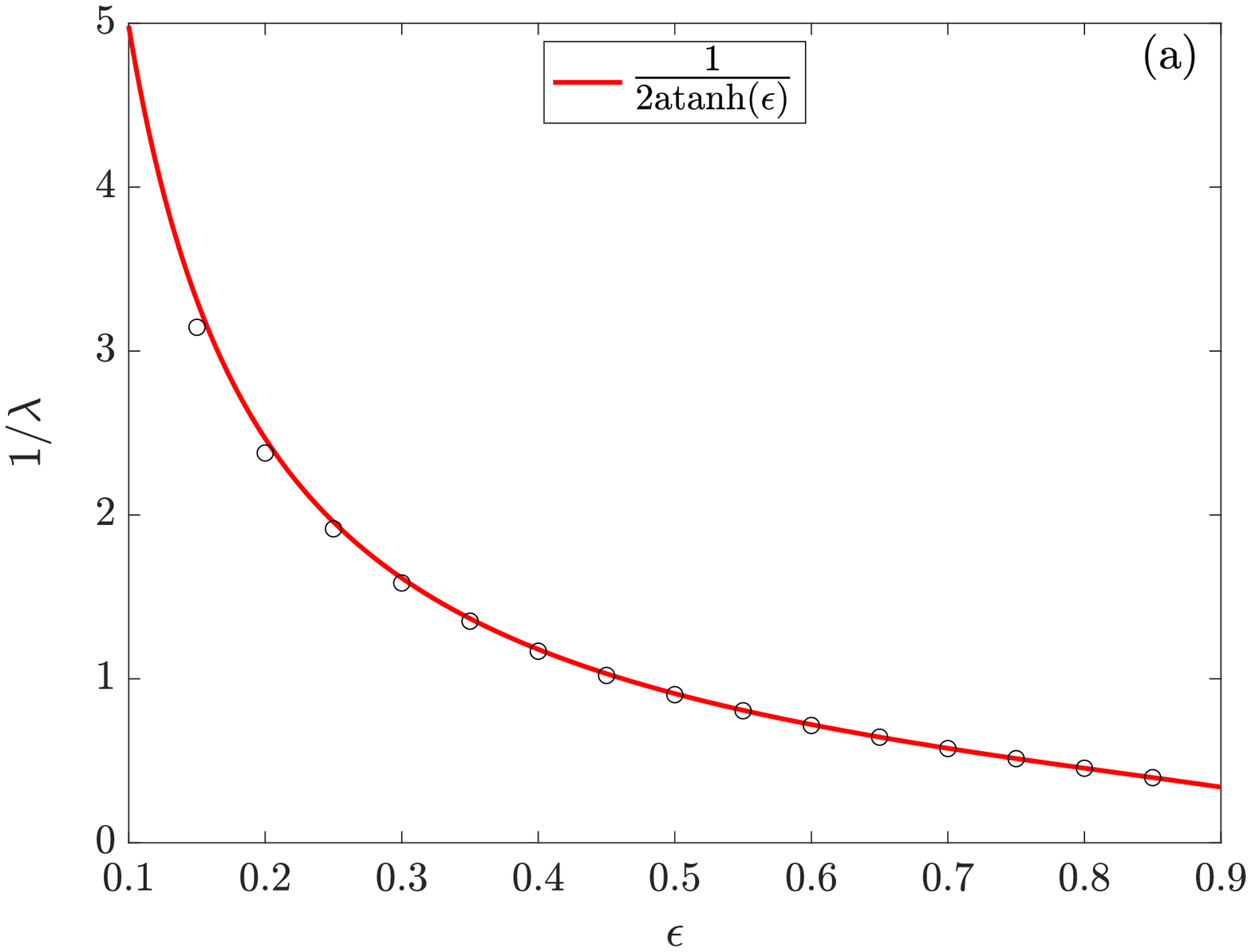}
	\includegraphics[width=0.48\textwidth]{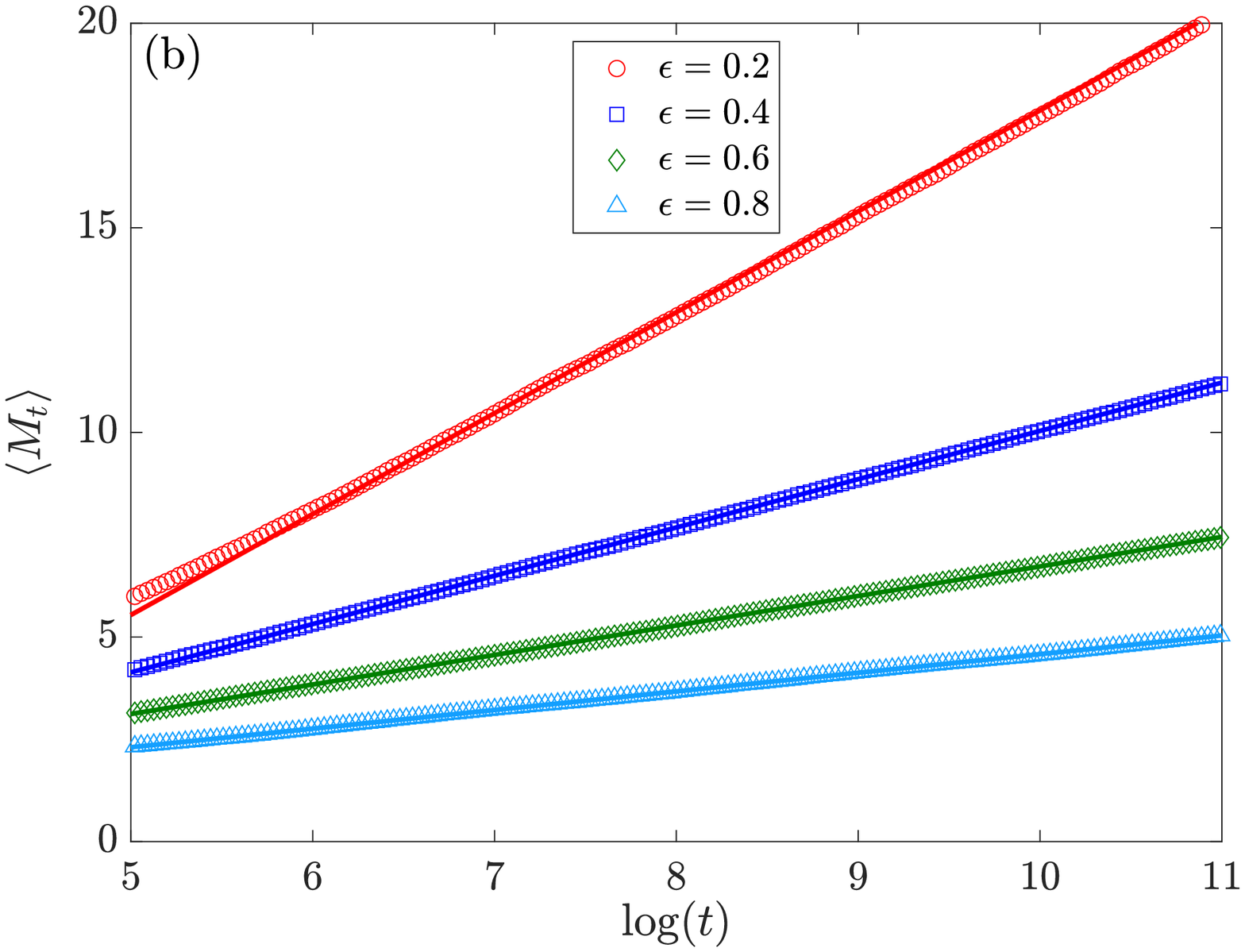}
	\caption{(a) Coefficients of the asymptotic logarithmic growth describing the expected maximum, see \eref{eq:beta0_Max}; the slopes have been obtained from a linear fit of $ \langle M_t\rangle $ versus $ \log(t) $, while the solid red curve represents the theoretical value. (b) Representation of the temporal behaviour of the expected maximum, as a function of $ \log(t) $; the growth is well described by the solid lines, which have been obtained by fixing the value of the slopes to $ 1/(2\mathrm{atanh}(\epsilon)) $ and leaving the intercepts as free parameters of the fit. All data in both panels have been obtained by evolving $ 10^4 $ walks up to $ 10^5 $ steps.}
	\label{fig: gillis beta0}
\end{figure}

\subsection{Logarithmic correction to the Gillis original drift.}
As a last example, we can think of adding just a weak logarithmic correction to the original drift (similarly to \cite{Englander2019}). As usual we consider $ \mathcal{R}(j)=1-\mathcal{L}(j) $ for any $ j $, and define
\begin{equation}
	2\mathcal{L}(j)=\begin{dcases}
		1\qquad&\mbox{if}\quad j=0\,,\\
		1+\frac{\epsilon}{j} \qquad&\mbox{if}\quad |j|=1\,,\\
		1+\frac \epsilon{j}\log\left(|j|\right)\qquad&\mbox{if}\quad |j|>1\,,
	\end{dcases}
\end{equation}
from which we observe that the first two moments of the increments are
\be
\mu_1(x)= -\frac\epsilon x \log(|x|)\quad\left(\mbox{for }|x|>1\right), \qquad\mu_2(x)=1.
\ee
Hence according to Lamperti criteria, see \eref{eq: lamp}, the walk is positive recurrent for $ \epsilon>0 $. In this range it is possible to recursively compute the stationary distribution
\begin{equation}
	\pi_j=\frac{\pi_0}{1+\epsilon}\cdot\begin{cases}
		1&\mbox{if}\quad |j|=1\\
		\displaystyle\frac{2(1-\epsilon)}{2+\epsilon\log\left(2\right)}&\mbox{if}\quad |j|=2\\
		\displaystyle\frac{|j|(1-\epsilon)}{|j|+\epsilon\log\left(|j|\right)} \prod_{k=2}^{|j|-1}\frac{k-\epsilon\log(k)}{k+\epsilon\log(k)}&\mbox{if}\quad |j|>2,
	\end{cases}
\end{equation}
and we observe that for $ |j|>2 $ we can write
\begin{equation}
	\pi_j=\pi_0\exp\left[-2\sum_{k=2}^{|j|-1}\mathrm{atanh}\left(\frac{\epsilon}{k}\log(k)\right)-\log\left(1+\frac{\epsilon\log(|j|)}{|j|}\right)-2\mathrm{atanh}(\epsilon)\right].
\end{equation}
With the same reasoning adopted for the supercritical model, we can finally conclude that the discrete random walk is asymptotically equivalent to a diffusing particle moving in a potential of the form
\be\label{eq:PotLog}
V(x)\sim\frac \epsilon 2 \log^2\left(\frac{|x|}\eta\right)\qquad \mbox{as}\quad x\to\pm\infty\,,
\ee
corresponding to a stationary solution whose tail decay as $x\to\pm\infty$
\be\label{eq: regLog}
\pi(x)\propto \rme^{-\epsilon\log^2(|x|)}
\ee
induces a trivial moments spectrum. For a complete characterization of the maximum statistics, as will soon become clear, we would need at least to properly determine the value of the asymptotic constant $\eta$ in \eqref{eq:PotLog}, in contrast to what happens in the presence of a power-law potential. However, we still have enough information to get the correct scaling with respect to time.

By computing the asymptotic behaviour of the integral
\be 
\int_0^x \, dy\, \rme^{\epsilon \log^2(y/\eta)}=\frac{\eta \rme^{-1/4\epsilon}}{2\sqrt{\epsilon}}\frac{\rme^w} {\sqrt{w}}+\dots\qquad \mbox{with}\quad w\coloneqq \left( \frac{2\epsilon\log(x/\eta)+1}{2\sqrt{\epsilon}}\right)^2\,,
\ee
we obtain again an equation of the form \eqref{eq: lambert} with coefficients
\be 
A=\rme\,,\qquad B=\frac {Ct\sqrt{\epsilon}} {\eta\langle \tau\rangle } \rme^{1/4\epsilon}\,,\qquad E=\frac 1 2 \,,
\ee
whose solution is
\be 
w=-\frac 1 2 W\left( -2\left( \frac {Ct\sqrt{\epsilon}} {\eta\langle \tau\rangle } \rme^{1/4\epsilon}\right)^{-2}\right)\,.
\ee
This time we have to refer to the secondary branch of the Lambert $W$ function and so we need to use the estimate
\be 
W(x)=-y-\log(y)+\dots \qquad \mbox{with}\quad y=\log\left( -\frac 1 x \right)\qquad \mbox {as} \quad x\to 0^-\,.
\ee
In conclusion, we find that for $t,x\gg1$
\be 
w\sim \log \left ( \frac{C\sqrt{\epsilon}\rme^{1/4\epsilon}}{\eta\sqrt{2}\langle \tau\rangle} t\right)\qquad \implies\qquad x\sim \eta \rme^{-1/2\epsilon }\rme^{\sqrt{\frac{\log(t)}{\epsilon}}}\,.
\ee

\subsection{Overview and comparison with random walks with stochastic resetting and ergodic Markov processes in confining potentials}\label{sec:overview}
Another useful technique to induce positive-recurrence is considering stochastic processes under resetting \cite{Evans2020}: by continually returning a symmetric random walk with i.i.d. jumps or a diffusing particle to its initial condition, one can generate an effective potential which drives the system to reach a nonequilibrium stationary state. Some preliminary investigations on the extreme value statistics have been recently implemented for the simplest case of Poissonian resetting \cite{Majumdar2022,Godreche2022}, where the position reset occurs randomly in time with a constant rate. For an arbitrary resetting probability, heuristic guidelines are available in order to determine the correct scaling of our interest: for at least exponential step length distributions, whose tail behavior is also inherited by the stationary distribution causing trivial constant moments, the expected maximum (as well as the mean value of the record number) follows a slow logarithmic growth in time. Actually, the same feature is shared more generally by any distribution in the limit of small resetting rates. 

Before going any further, let us stress the similarity with the results of the supercritical Gillis model, as anticipated in the previous subsection. In our context, the pairing between the stationary distribution decay and the expected maximum growth becomes clear in \eqref{eq: max}. 
Nevertheless, one can notice a richer behaviour in the limiting distribution of the maximum for random walks with resetting: in the presence of a weak resetting, by varying the mean number of resettings, the limiting law for the properly rescaled maximum interpolates between a half-gaussian and a Gumbel distribution, whereas we directly recover the latter one in our results.
 
\subsubsection{Non-trivial moments spectrum.}
For subexponential jump distributions, instead, a considerable difference arises. First of all, we notice that in analogy with the standard Gillis model in the ergodic regime, a power-law decay of the stationary distribution, induced by the jump distribution, ensures not completely trivial transport properties, maintaining a time dependence of some moments. In this context, the expected maximum grows faster than a logarithm, as a power of the number of steps, in contrast to the mean record number. In particular, for L\'evy flights with finite mean the characteristic exponent remains unchanged regardless of the presence of resetting events: there is a change only in the moments spectrum due to the asymptotic properties of the stationary distribution. For more details, refer to \cite{Godreche2022}.

\subsubsection{Lyapunov functions and spectral interlacing.}
In \cite{Hartich2019a,Hartich2019b} the authors further deepen the duality between first-passage times and extreme value statistics but from the point of view of spectral analysis. For this reason, they are forced to consider strongly confining potentials which not only guarantee the existence of a stationary distribution, but also a spectral expansion of the Fokker-Planck operator with discrete eigenvalues. This technique clearly can not be applied to less binding potentials, for instance the logarithmic one, where a continuous spectrum starting at zero appears (see also \cite{dechant}).

By way of illustration, for an Ornstein-Uhlenbeck process, that is a harmonic potential, it is easy to recover (in a similar way to subsection \ref{subsec:GGL}) the scaling behavior $\sqrt{\log(t)}$ and the Gumbel distribution, as stated by the authors by means of the spectral analysis, see \cite{Hartich2019b}.

%
%
%
%
%
\section{Conclusions}\label{sec:conclusion}
In this work we have analysed the statistics of the maximum for different examples of positive recurrent random walks stemming from the Gillis model, which represents one of the few exactly solvable non-homogeneous random walks known in the literature. 

By tuning the parameter of the position-dependent drift in such a way that the existence of a stationary probability distribution is ensured, non-trivial transport and statistical properties arise. In particular, in the previous sections we emphasized the relation between the tail decay of the stationary measure and the limiting distribution of the maximum, and we compared it to analogous results recently obtained in other contexts.

Yet many unresolved questions about the effects of the lack of homogeneity remain open and we hope that further interesting results will be developed in a near future.

\section*{Acknowledgements}
 The authors are partially supported by  the PRIN Grant 2017S35EHN \emph{``Regular and stochastic behavior in dynamical systems''} (MIUR, Italy). R.A. acknowledges an association to the GNFM group of INDAM.

\appendix
\section{Gillis potential and regularization at the origin}\label{app: gillis}

\begin{figure}
	\centering
	\includegraphics[width=0.6\linewidth]{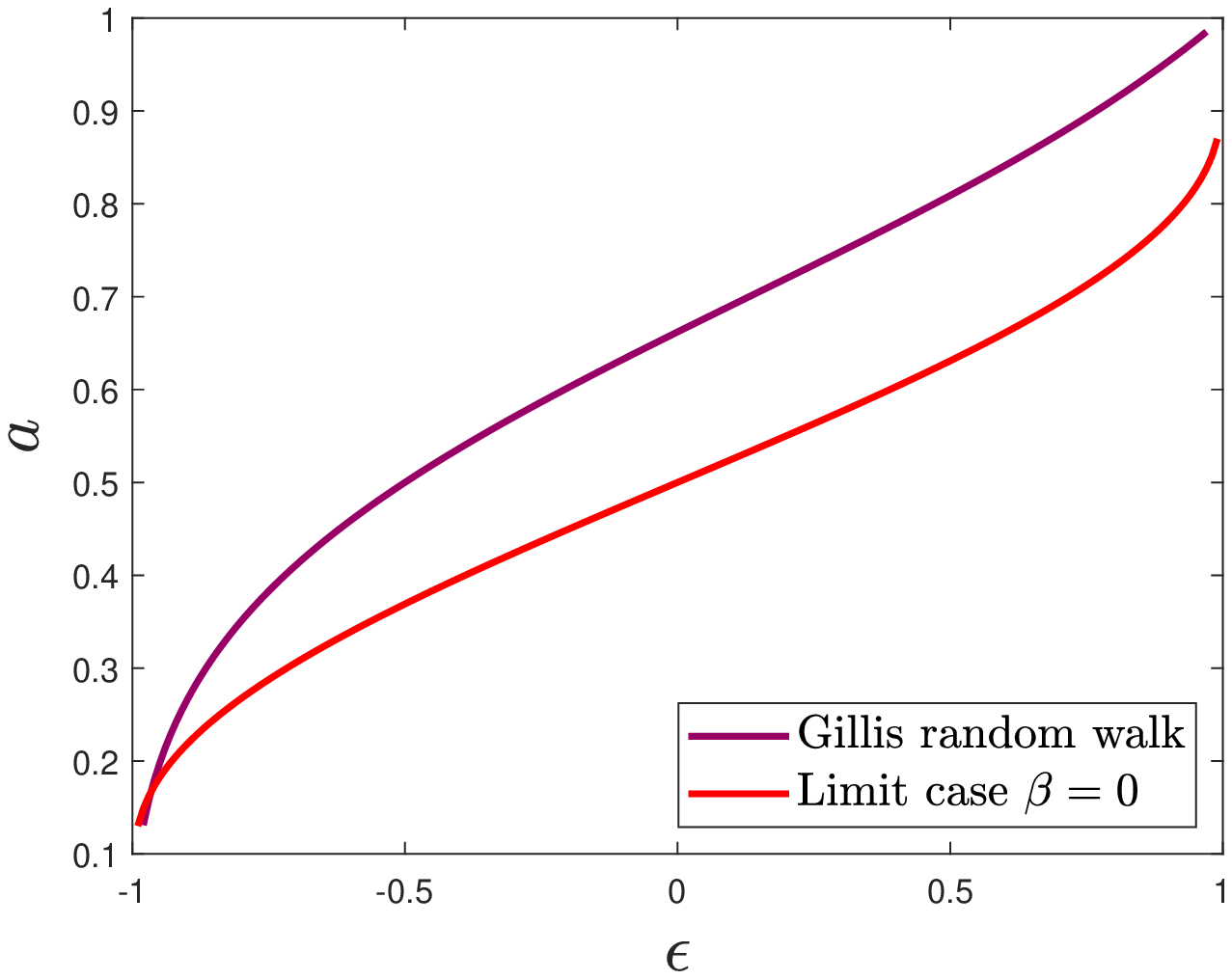}
	\caption{Plot of the size $ a $ versus $ \epsilon $, defined as a solution of \eqref{eq:a_sol} for the Gillis random walk and of \eqref{eq:a_beta_0} for the limit case $\beta=0$ of the supercritical model.}
	\label{fig:a_sol}
\end{figure}

The complete form of the potential of the Gillis random walk can be directly derived from the stationary distribution $ \pi_j\, $, see \eqref{eq: stat}. The latter is defined on $\mathbbm{Z}$, but can be easily extended to a setting on the real line, by expressing the Pochhammer symbol in terms of the Gamma function \cite{Abr-Steg}
\begin{equation}
	(x)_n\equiv\frac{\Gamma(x+n)}{\Gamma(x)}\,.
\end{equation}
Hence, for $ x\neq 0 $, we can define
\begin{equation}
	\pi(x)=\frac{2\epsilon-1}{2\epsilon}\frac{\Gamma(1+\epsilon)}{\Gamma(1-\epsilon)}\frac{\Gamma(|x|-\epsilon)}{\Gamma(|x|+\epsilon+1)}|x|\,,
\end{equation}
which is the continuum equivalent of $ \pi_j $ for $ j\neq 0 $. Ideally, for $ x=0 $ we want $ \pi(0)=(2\epsilon-1)/2\epsilon $, but the previous expression converges instead to $ 0 $. Therefore we need to introduce a regularizing region around the origin, of size $ 2a $, in such a way that $ \pi(x) $ is continuous and
\begin{equation}
	\pi(x)=
	\begin{dcases}
		\frac{2\epsilon-1}{2\epsilon} &\quad\text{ if }\:|x|<a\,,\\
		\frac{2\epsilon-1}{2\epsilon}\frac{\Gamma(1+\epsilon)}{\Gamma(1-\epsilon)}\frac{\Gamma(|x|-\epsilon)}{\Gamma(|x|+\epsilon+1)}|x| &\quad\text{ if }\:|x|>a.
	\end{dcases}
\end{equation} 
A proper choice for $ a $ would be to take it as the solution of
\begin{equation}\label{eq:a_sol}
	\frac{\Gamma(1+\epsilon)}{\Gamma(1-\epsilon)}\frac{\Gamma(|x|-\epsilon)}{\Gamma(|x|+\epsilon+1)}|x|=1.
\end{equation}
Note that this equation can be solved numerically not only for $ \epsilon>\tfrac 12 $, i.e., in the ergodic regime, but also for $ \epsilon<\tfrac 12 \,$, see figure \ref{fig:a_sol}. Hence this particular choice is valid for both regimes.

\begin{figure}
	\centering
	\includegraphics[width=0.6\linewidth]{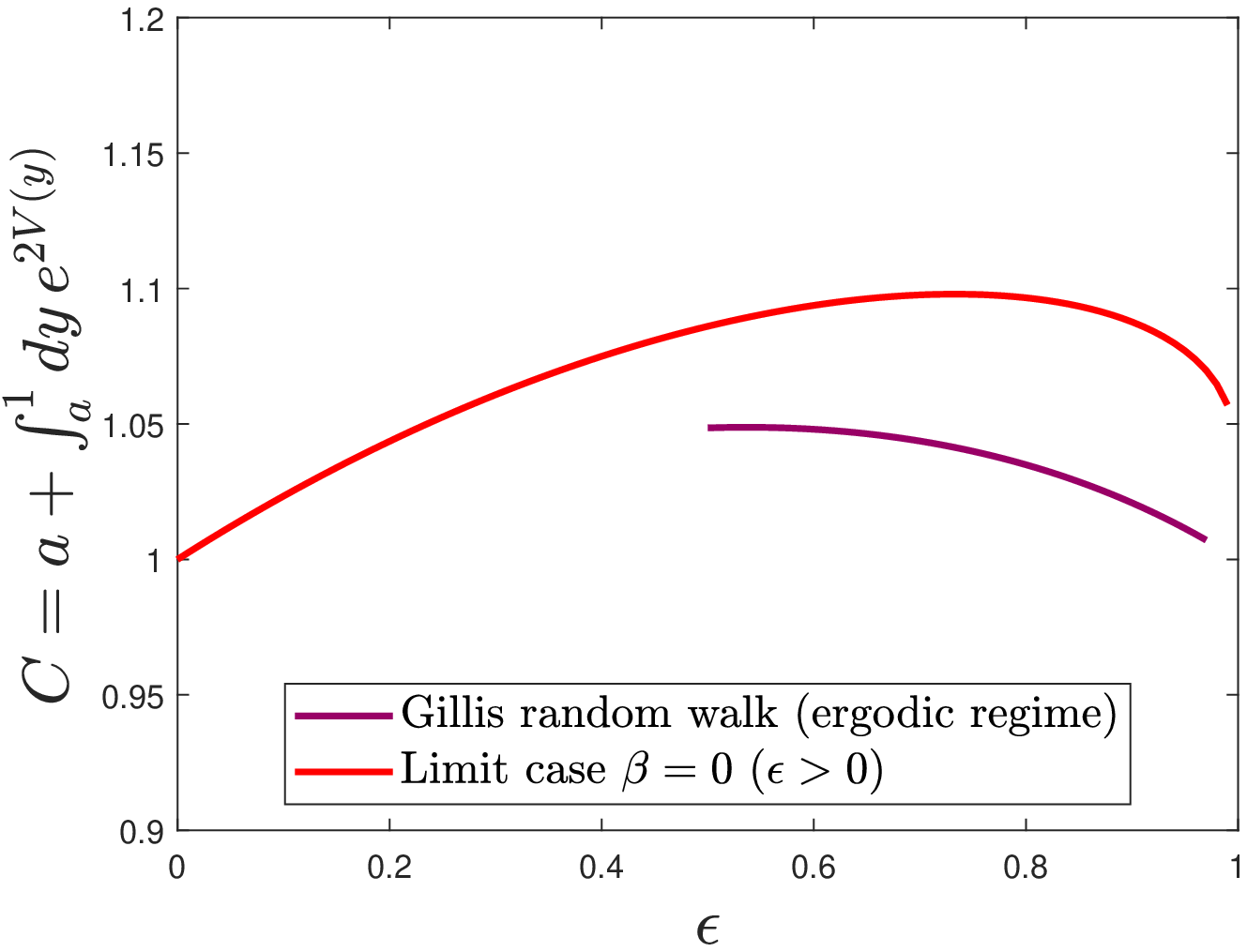}
	\caption{Plot of the constant $C$, defined in \eqref{eq: C} for the Gillis random walk and in \eqref{eq: C_beta_0} for the limit case $\beta=0$ of the supercritical model, versus $ \epsilon $.}
	\label{fig: C}
\end{figure}

Moreover, since the potential is related to the stationary distribution by means of \eqref{eq: statDist}, that is
\begin{equation}
	\pi(x)=\frac{\rme^{-2V(x)}}{Z}\,,
\end{equation}
where $ 1/Z\coloneqq(2\epsilon-1)/2\epsilon $ is the normalization constant, we can therefore define
\begin{equation}
	V(x)=
	\begin{dcases}
		0 &\quad \text{for }\:|x|<a\,,\\
		\frac 1 2\log\left[\frac{\Gamma(1-\epsilon)}{\Gamma(1+\epsilon)}\frac{\Gamma(|x|+1+\epsilon)}{|x|\Gamma(|x|-\epsilon)}\right] &\quad \text{for }\:|x|>a\,.
	\end{dcases}
\end{equation}
The potential is indeed of the form $ V(x)\sim \epsilon\log|x| $ for large $ |x| $, due to the asymptotic decay of $ \pi(x) $, but it is identified by a different expression for smaller values of $ |x| $. In conclusion, \eqref{eq: pot} is a good approximation for long-range applications, whereas a more careful analysis would be needed for intermediate values of $|x|$. 
In particular, notice that in \eqref{eq: Q} we have to set $\bar x=1$ as explained in the main text, as a consequence we find that the constant $C$ in \eqref{eq: max} is given by
\be \label{eq: C}
C=a+\int_a^1 \frac{\Gamma(1-\epsilon)\Gamma(1+\epsilon+y)}{\Gamma(1+\epsilon) y \Gamma(y-\epsilon)}\, dy\approx 1\,,
\ee
as can be seen in figure \ref{fig: C}.

Similar computations also apply to the limit case $\beta=0$ of the supercritical model, where we explicitly wrote the discrete stationary distribution \eqref{eq:beta0Stat}. Thus, we can define the continuum potential
\begin{equation}
	V(x)=
	\begin{dcases}
		0 &\quad \text{for }\:|x|<a\,,\\
		\frac 1 2\log\left[(1-\epsilon)\left(\frac{1+\epsilon}{1-\epsilon}\right)^{|x|}\right] &\quad \text{for }\:|x|>a\,,
	\end{dcases}
\end{equation}
with $a$ taken as the solution of
\be\label{eq:a_beta_0}
\frac 1 {1-\epsilon} \left( \frac{1-\epsilon}{1+\epsilon}\right)^{|x|}=1\,,
\ee
and consequently
\be\label{eq: C_beta_0}
C=a+\frac{1-\epsilon}{\log\left(\frac{1+\epsilon}{1-\epsilon}\right)}\left[\frac{1+\epsilon}{1-\epsilon}-\left(\frac{1+\epsilon}{1-\epsilon}\right)^a\right]\,.
\ee
See figures \ref{fig:a_sol} and \ref{fig: C} for a numerical result.

\section{Continuum limit of the Gillis persistent random walk}\label{app: gillis pers}
In order to obtain the diffusion equation for the probability density function $p(x,t)$ corresponding to a random walk defined on $\mathbbm{Z}$, first of all we have to fix some notation. We set the parameter equal to $\epsilon_-$ when the particle moves toward the origin, $\epsilon_+$ if it gets away. Moreover, we denote
\begin{align*}
p_+(j,n)&:=\mathbb{P}\{\mbox{ the particle is at site $j$ after $n$ steps and goes right }\},\\
p_-(j,n)&:=\mathbb{P}\{\mbox{ the particle is at site $j$ after $n$ steps and goes left }\},\\
t_j^+,r_j^+&:= \mbox{ transmission and reflection coefficients at site $j$ in the outward direction},\\
t_j^-,r_j^-&:= \mbox{ transmission and reflection coefficients at site $j$ in the inward direction},
\end{align*}
where
$$
t_j^\pm=\frac{1}{2}\left(1-\frac{\epsilon_\pm}{|j|}\right)\,, \qquad r_j^\pm=\frac{1}{2}\left(1+\frac{\epsilon_\pm}{|j|}\right)\,.
$$
We immediately notice that we have to define the master equation separately on the positive and negative integers. For instance, if $j\in\mathbb{Z}^+$
\begin{equation*}
\begin{cases}
p_+(j,n+1)=t_j^+p_+(j-1,n)+r_j^-p_-(j+1,n)\,,\\
p_-(j,n+1)=t_j^-p_-(j+1,n)+r_j^+p_+(j-1,n)\,.\\
\end{cases}
\end{equation*}
By summing up the two equations, we get 
\begin{align*}
p(j,n+1)&=p_+(j,n+1)+p_-(j,n+1)=(t_j^++r_j^+)p_+(j-1,n)+(t_j^-+r_j^-)p_-(j+1,n)\\
&=p_+(j-1,n)+p_-(j+1,n)\,.
\end{align*}
Now, by introducing the definitions $x:= j\delta x$, $t:=n\delta t$ ($\delta x, \delta t\to 0$), the probability density $p(x,t)$ such that $p(j,n)=\delta x p(x,t)$ and expanding up to the second and first order in $\delta x$, $\delta t$ respectively, we arrive at the following diffusive approximation
$$
\dot{p}(x,t)=-m'(x,t)+\frac{1}{2}\frac{\delta x^2}{\delta t}p''(x,t)\,,
$$
where $m(j,n)\coloneqq p_+(j,n)-p_-(j,n)=\delta t \, m(x,t)=\delta x[p_+(x,t)-p_-(x,t)]$ is the probability current, $\frac{\delta x^2}{\delta t}\to D_0(=1)$ as $\delta x,\delta t\to 0$ and the single dot represents the time derivative, whereas the prime notation is used for spatial derivatives.\\
At this point, we have to consider also the difference in order to obtain the second coupled equation, that is
$$m(j,n+1)=(1-2r_j^+)p_+(j-1,n)-(1-2r_j^-)p_-(j+1,n)\,.$$
We observe that, in the continuum limit, by dropping higher order terms as before, we get
\begin{align*}
\delta t m(x,t)=\delta t m(x,t)&-\delta x^2p'(x,t)-\delta t(r_+(x)+r_-(x))m(x,t)\\
&\qquad -\delta x(r_+(x)-r_-(x))p(x,t)+\delta x^2(r_+(x)+r_-(x))p'(x,t)\,,
\end{align*}
since 
$$2(r_+p_+\pm r_-p_-)=(r_++r_-)(p_+\pm p_-)+(r_+-r_-)(p_+\mp p_-)\,.$$
Hence we can write
\be\label{eq: appCurr}
m(x,t)=-\frac{\epsilon_+-\epsilon_-}{2x}p(x,t)\,,
\ee
given that 
$$r_++r_-=1+\frac{\epsilon_+-\epsilon_-}{2x}\delta x\,,\qquad r_+-r_-=\frac{\epsilon_+-\epsilon_-}{2x}\delta x\,.$$
 In conclusion, we find again the diffusion equation for a particle in a logarithmic potential
\begin{equation*}
\frac{\partial}{\partial t} p=\frac1 2 \left[\frac{\partial ^2}{\partial x^2}+(\epsilon_+-\epsilon_-)\frac{\partial}{\partial x}\left(\frac{1}{x}p\right)\right],
\end{equation*}
which is the same of a standard Gillis random walk: it is sufficient to set $\epsilon=\frac{\epsilon_+-\epsilon_-}{2}$.

For the sake of completeness, on the negative integers, instead, the roles of $t_+$ and $t_-$ are reversed
\begin{equation*}
\begin{cases}
p_+(j,n+1)=t_j^-p_+(j-1,n)+r_j^+p_-(j+1,n)\,,\\
p_-(j,n+1)=t_j^+p_-(j+1,n)+r_j^-p_+(j-1,n)\,.\\
\end{cases}
\end{equation*}
The sum remains unchanged, whereas for the difference we have to replace the following relations
$$2(r_-p_+\pm r_+p_-)=(r_-+r_+)(p_+\pm p_-)+(r_--r_+)(p_+\mp p_-)\qquad \mbox{and}\qquad r_--r_+=\frac{\epsilon_--\epsilon_+}{2|x|}\delta x\,.$$
In the end, we still obtain
$$m(x,t)=-\frac{\epsilon_+-\epsilon_-}{2x}p(x,t) \qquad\mbox{for}\quad x<0\,.$$ 
As a final comment, observe that for a generalized drift identified by \eqref{eq: persGen} the significant quantities for the continuum limit are $r_++r_-=2(1-q)+\frac{\epsilon_+-\epsilon_-}{2x}\delta x$ and $r_+-r_-=\frac{\epsilon_+-\epsilon_-}{2x}\delta x$. As a consequence, the generalized \eqref{eq: appCurr} reads
$$m(x,t)=\frac{1-2q}{2(1-q)}D_0p'(x,t)-\frac{D_0}{2(1-q)}\frac{\epsilon_+-\epsilon_-}{2x}p(x,t)\,,$$
 and the related diffusion equation is therefore \eqref{eq: persGenCL}.

A rigorous analysis of recurrence properties for these generalized Lamperti drifts can be found in \cite{wade}, where Lamperti criteria are extended by means of Theorem $2.6$: according to their notation, it is sufficient to take $q_{++}=q_{--}=q$, $q_{+-}=q_{-+}=1-q$, $\gamma_{++}=-\gamma_{+-}=-\frac{\epsilon_+}{2}$, $\gamma_{--}=-\gamma_{-+}=-\frac{\epsilon_-}{2}$, $d_+=-d_-=2q-1$, $e_+=-\epsilon_+$, $e_-=\epsilon_-$, $t_i^2=1$, $d_{++}=-d_{--}=q$, $d_{+-}=-d_{-+}=q-1$, $\pi_{\pm}=\frac{1}{2}$, $a=\left(0,\frac{1-2q}{1-q}\right)$ in order to go back to our examples. 

\section{Tail decay of the stationary distribution for the supercritical Gillis random walk}\label{app:StatSuperGillis}
\begin{figure}
	\centering
	\includegraphics[width=0.495\linewidth]{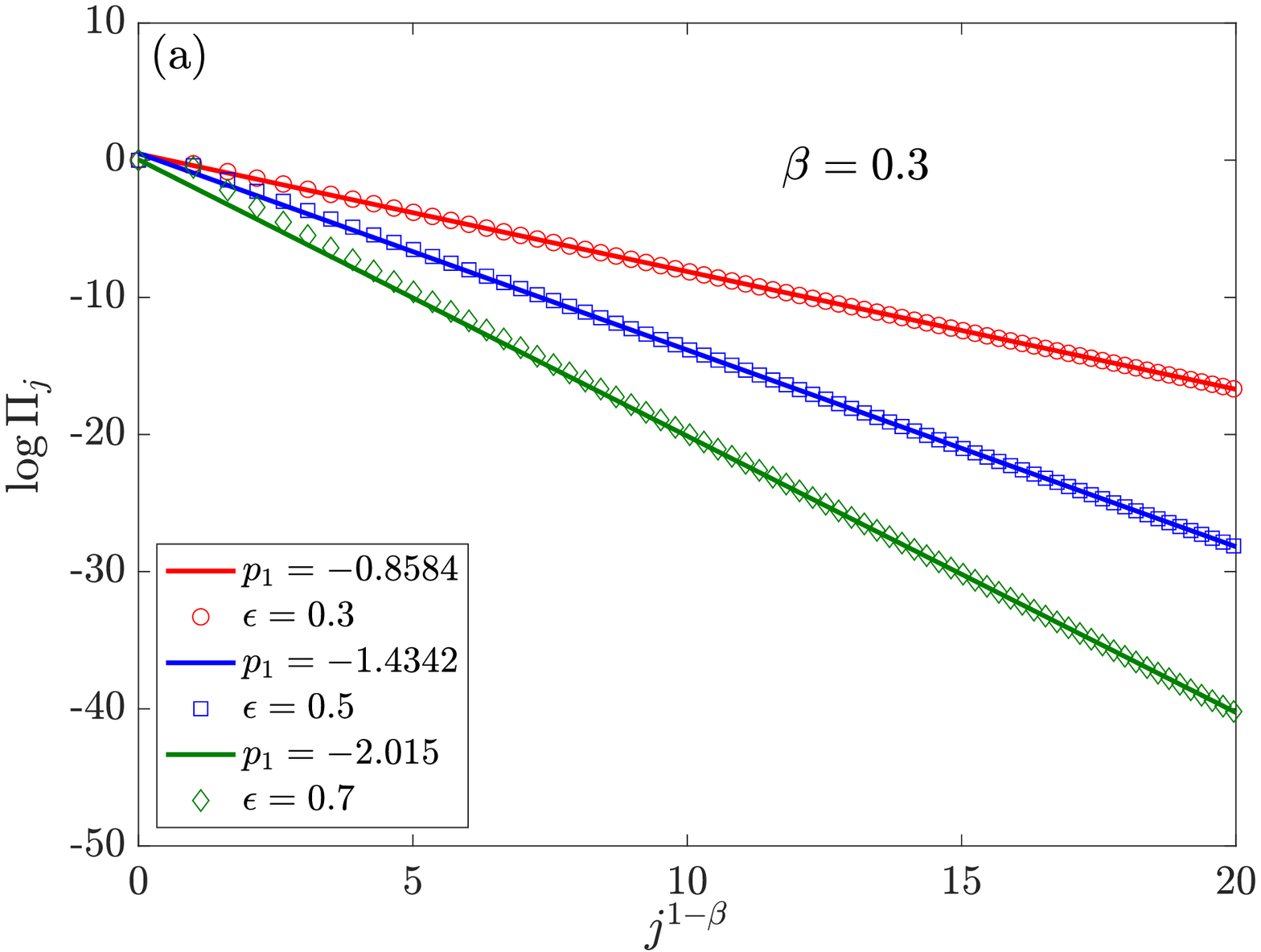}
	\includegraphics[width=0.495\linewidth]{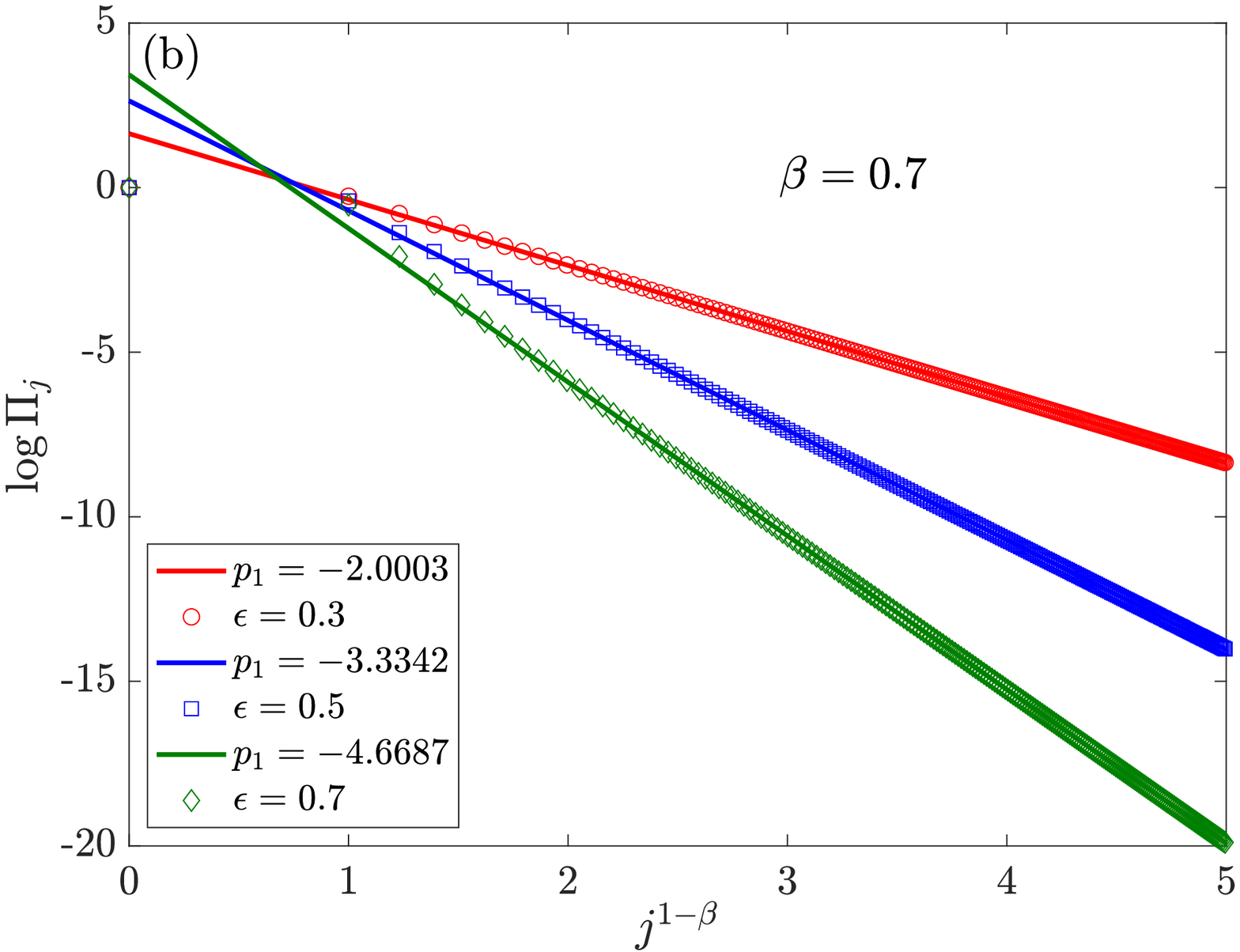}
	\caption{Gillis supercritical model: linear fit of $ \log\Pi_j $ versus $ j^{1-\beta} $, for $ j>0 $. Markers represent the exact values of $ \Pi_j $ obtained from \eref{eq:Pi_j}. The slope of each line is in perfect agreement with the theoretical prediction $ p_{\mathrm{th}}=-2\epsilon/(1-\beta) $.}
	\label{fig:Appendix_super_Pot}
\end{figure}
In order to substantiate the approximation \eqref{eq: statApprox} in the main text, one has to exclude the existence of relevant second-order corrections. 
Getting rid of the unknown normalization in \eqref{eq: superStat}, we can write
\be\label{eq:Pi_j}
\Pi_j\coloneqq\frac{\pi_j}{\pi_0}=\begin{cases}
\displaystyle\frac 1 {1+\epsilon}\qquad &\mbox{if}\quad |j|=1\,,\\
\displaystyle\frac{|j|^\beta}{|j|^\beta+\epsilon}\prod_{k=1}^{|j|-1}\frac{|k|^\beta-\epsilon}{|k|^\beta+\epsilon}\qquad&\mbox{if}\quad |j|>1\,,\\
\end{cases}
\ee
with $\Pi_0=1\,$. In accordance with \eqref{eq: statApprox}, we perform a numerical check by looking for a stretched-exponential law
\be
\Pi_j=p_2e^{p_1j^{1-\beta}}\,,
\ee
which equivalently reads
\be
\log(\Pi_j)=p_1j^{1-\beta}+\log(p_2)\,.
\ee
As can be seen in figure \ref{fig:Appendix_super_Pot}, numerical simulations confirm this linear behaviour with $p_1=-\frac{2\epsilon}{1-\beta}$. 

\section*{References}
\bibliographystyle{iopart-num}

\providecommand{\newblock}{}

\end{document}